\documentclass[sigconf]{acmart}
\AtBeginDocument{%
  \providecommand\BibTeX{{%
    \normalfont B\kern-0.5em{\scshape i\kern-0.25em b}\kern-0.8em\TeX}}}

\copyrightyear{2024} 
\acmYear{2024} 
\setcopyright{acmlicensed}\acmConference[CHI '24]{Proceedings of the CHI Conference on Human Factors in Computing Systems}{May 11--16, 2024}{Honolulu, HI, USA}
\acmBooktitle{Proceedings of the CHI Conference on Human Factors in Computing Systems (CHI '24), May 11--16, 2024, Honolulu, HI, USA}
\acmDOI{10.1145/3613904.3641901}
\acmISBN{979-8-4007-0330-0/24/05}

\newcommand{\revision}[1]{\textcolor{black}{#1}} 
\newcommand{\cm}[1]{\textcolor{black}{#1}} 
\usepackage{ulem}

\usepackage[utf8]{inputenc}
\DeclareUnicodeCharacter{02BB}{'}

\usepackage[most]{tcolorbox}
\definecolor{cback}{HTML}{EDEFF1}
\definecolor{cframe}{HTML}{B9C4CA}
\definecolor{cgrey}{HTML}{666666}
\definecolor{cdeeppurple}{HTML}{673AB7}
\definecolor{cvividorange}{HTML}{FF5722}
\definecolor{cforestgreen}{HTML}{228B22}
\definecolor{cmidnightblue}{HTML}{191970}

\newtcbox{\inlinebox}[1][]{enhanced,
 box align=base,
 nobeforeafter,
 colback=cback,
 colframe=cframe,
 size=small,
 left=0pt,
 right=0pt,
 boxsep=2pt,
 #1}
 
 \newtcbox{\inlinethinbox}[1][]{enhanced,
 box align=base,
 nobeforeafter,
 colback=cback,
 colframe=cframe,
 size=small,
 left=0pt,
 right=0pt,
 boxsep=1pt,
 #1}
 
\newtcbox{\nnboxx}[1][]{enhanced,
 box align=base,
 nobeforeafter,
 colback=cgrey,
 colframe=cgrey,
 colupper=white,
 size=small,
 left=1pt,
 right=1pt,
 boxsep=0.2mm,
 arc = 0mm,
 outer arc=0mm,
 #1}

\newtcbox{\nnboxy}[1][]{enhanced,
 box align=base,
 nobeforeafter,
 colback=cvividorange,
 colframe=cgrey,
 colupper=white,
 size=small,
 left=1pt,
 right=1pt,
 boxsep=0.2mm,
 arc = 0mm,
 outer arc=0mm,
 #1}

\newtcbox{\nnboxz}[1][]{enhanced,
 box align=base,
 nobeforeafter,
 colback=cforestgreen,
 colframe=cgrey,
 colupper=white,
 size=small,
 left=1pt,
 right=1pt,
 boxsep=0.2mm,
 arc = 0mm,
 outer arc=0mm,
 #1}

\newtcbox{\nnboxo}[1][]{enhanced,
 box align=base,
 nobeforeafter,
 colback=cmidnightblue,
 colframe=cgrey,
 colupper=white,
 size=small,
 left=1pt,
 right=1pt,
 boxsep=0.2mm,
 arc = 0mm,
 outer arc=0mm,
 #1}

\begin{document}

\title[Exploring the Opportunity of Augmented Reality (AR)]{Exploring the Opportunity of Augmented Reality (AR) in Supporting Older Adults Explore and Learn Smartphone Applications}


\author{Xiaofu Jin}
\orcid{0000-0003-3837-2638}
\affiliation{
  \institution{The Hong Kong University of Science and Technology}
  \city{Hong Kong SAR}
  \country{China}
}
\email{xjinao@connect.ust.hk}

\author{Wai Tong}
\orcid{0000-0001-9235-6095}
\affiliation{
  \institution{Texas A\&M University}
  \city{Texas}
  \country{USA}
}
\email{wtong@tamu.edu}

\author{Xiaoying Wei}
\orcid{0000-0003-3837-2638}
\affiliation{
  \institution{The Hong Kong University of Science and Technology}
  \city{Hong Kong SAR}
  \country{China}
}
\email{xweias@connect.ust.hk}

\author{Xian Wang}
\orcid{0000-0003-1023-636X}
\affiliation{
  \institution{The Hong Kong Polytechnic University}
  \city{Hong Kong SAR}
  \country{China}
}
\email{xian0203.wang@polyu.edu.hk}

\author{Emily Kuang}
\orcid{0000-0003-4635-0703}
\affiliation{
  \institution{Rochester Institute of Technology}
  \city{Rochester}
  \country{USA}
}
\email{ek8093@rit.edu}

\author{Xiaoyu Mo}
\orcid{0000-0001-6009-6487}
\affiliation{
  \institution{The Hong Kong University of Science and Technology}
  \city{Hong Kong SAR}
  \country{China}
}
\email{xmoac@connect.ust.hk}

\author{Huamin Qu}
\orcid{0000-0002-3344-9694}
\affiliation{
  \institution{The Hong Kong University of Science and Technology}
  \city{Hong Kong SAR}
  \country{China}
}
\email{huamin@cse.ust.hk}

\author{Mingming Fan}
\authornote{Corresponding Author}
\orcid{0000-0002-0356-4712}
\affiliation{
  \institution{The Hong Kong University of Science and Technology (Guangzhou)}
  \city{Guangzhou}
  \country{China}
}
\affiliation{
  \institution{The Hong Kong University of Science and Technology}
  \city{Hong Kong SAR}
  \country{China}
}
\email{mingmingfan@ust.hk}


\begin{abstract}

The global aging trend compels older adults to navigate the evolving digital landscape, presenting a substantial challenge in mastering smartphone applications. While Augmented Reality (AR) holds promise for enhancing learning and user experience, its role in aiding older adults' smartphone app exploration remains insufficiently explored. Therefore, we conducted a two-phase study: (1) a workshop with 18 older adults to identify app exploration challenges and potential AR interventions, and (2) tech-probe participatory design sessions with 15 participants to co-create AR support tools. Our research highlights AR's effectiveness in reducing physical and cognitive strain among older adults during app exploration, especially during multi-app usage and the trial-and-error learning process. We also examined their interactional experiences with AR, yielding design considerations on tailoring AR tools for smartphone app exploration. Ultimately, our study unveils the prospective landscape of AR in supporting the older demographic, both presently and in future scenarios.

\end{abstract}

\ccsdesc[500]{Human-centered computing~Empirical studies in HCI}
\ccsdesc[500]{Human-centered computing~Mixed / augmented reality}
\keywords{augmented reality, older adults, smartphone exploration, independent learning}



\maketitle

\section{Introduction}
Modern society has seen a rapid evolution in the adoption of digital technologies, and online platforms have become the primary way for people to access information and services. This shift has been so significant that some services are gradually replaced by or even offered exclusively through digital channels including online education courses~\cite{coursera, edx, udacity}, e-government services~\cite{SHAREEF201117}, and digital banking services~\cite{EY2020}. Despite the convenience of these digital solutions, individuals lacking technical expertise may face challenges when using them. This is particularly true for many older adults, who are impacted by various factors, such as cohort effects~\cite{Yusif2016Older}, the digital divide~\cite{ordonez2011elderly, Choudrive2018}, and age-related cognitive and physical declines~\cite{czaja2006factors, Mitzner2018Technology}. These factors create obstacles for older adults when it comes to learning how to use the complex features of smartphones. Consequently, many older adults use smartphones as feature phones, limiting their usage to basic functions such as making phone calls and sending text messages~\cite{BerenguerAnabela2017ASUA} and about a quarter of older adults who aged 65 and more still do not use their smartphone to access internet~\cite{pew2022techusers}.
Nevertheless, older adults are motivated to learn smartphones due to factors such as a desire to reduce loneliness~\cite{sims2017information}, perceived usefulness~\cite{o2008understanding, Venkatesh2012UTAUT2, Jin22I}, and social influence~\cite{Lin2020Exploring, Jin22I}. In addition, learning has been found to have a positive impact on various aspects of health for older adults (e.g., mental cognitive health, psychological aspects, and social aspects of health)~\cite{Tam2016Ageing, Hardy2017Baby, Aberg2016Nonformal}.

To support older adults in learning digital technologies, researchers have conducted studies to understand their learning needs and preferences. Older adults are found to be diverse and prefer more flexible support~\cite{Jin22I, Pang2021Technology}. In addition to the traditional support of training materials (e.g., instructions, videos) and video calls, researchers have explored different technologies such as e-learning platforms~\cite{kopeclivinglab2017}, social support applications ~\cite{mendelmeerkat2022} and personalized interactive guidance with trial-and-error support~\cite{jinsynapse2022}. 

Augmented Reality (AR) has demonstrated the potential to enhance user experience and improve learning outcomes. Researchers have leveraged AR to visualize challenging concepts and support real-world simulations with interactive objects, thus increasing content understanding, learning motivation, and long-term memory retention~\cite{Santos2014Augmented, albrecht2013effects, Khan2018Mathland, Radu2019What}. Such pioneering investigations suggest that AR may be able to help older adults learn smartphone technologies. Particularly, older adults generally prefer bigger screens~\cite{pandyataptag2018, Pang2021Technology}. AR can expand the display of supporting information beyond the limitation of physical screens, creating a more flexible and dynamic experience, and allowing for engaging with content from multiple spatial and visual angles. 
In addition, AR technology can integrate a greater amount of information within the 3D augmented space, which has the potential to reduce the clutter of visual information and thus provide better support for older adults. 

Rather than focusing solely on device usability, positive technology, which aims to enhance the overall quality of life by promoting human well-being, satisfaction, and contentment, is a potential alternative for older adults~\cite{grossi2020positive}. 
AR technology offers innovative ways of interacting with the world, which can help older adults enjoy the learning process in a more engaging, fun, and motivating manner~\cite{Mostajeran2020Augmented}. However, older adults are largely neglected in current AR research, with limited work focusing on exploring the potential of AR for improving older adults' well-being (e.g., enhance fall prevention of older adults~\cite{Bianco2016Augmented}, coach for balance training~\cite{Mostajeran2020Augmented}). Although usability issues still exist in current AR technologies, older adults expressed their acceptance of AR and found AR systems encouraging~\cite{Mostajeran2020Augmented}. This highlights an opportunity to leverage AR to support older adults in their learning process. \revision{This work primarily focuses on facilitating independent exploring and learning among older adults using AR. While social interaction have benefits in their learning, our emphasis is on how AR can empower them to autonomously learn and navigate digital technologies, addressing a key aspect of digital literacy. Nevertheless, the design principles for AR to effectively assist in the mobile technology exploring and learning process for older adults remain largely unexplored.} To fill in this gap, we explore the following research questions (RQs):
\begin{itemize} 
    \item RQ1: How could AR be \revision{potentially} leveraged to support older adults in exploring and learning smartphone apps \revision{independently}?
    \item RQ2: How might older adults want to interact with AR when exploring and learning smartphone apps?
\end{itemize}


  


Prior work has shown that technology probe-based participatory design could help identify opportunities of emerging technology for older adults, such as the future of the Internet of things (IoT) technologies~\cite{Pradhan2020Understanding}, and VR for intergenerational communication~\cite{Xiaoying2023Bridging}. Informed by this line of work, we took a similar approach by conducting a two-phase study: (1) A workshop on investigating what challenges older adults face most when learning smartphones and where AR could be applied to offer support. (2) Technology probe-based participatory design sessions to brainstorm and develop an interactive AR support tool collaboratively. Our findings illuminate the multifaceted advantages of AR, notably in alleviating physical and cognitive strains, especially during tasks like multi-app navigation and the nuanced trial-and-error learning processes. By deep-diving into the interactional experiences of older adults with AR, we derived specific design considerations that can shape the development of AR tools tailored to this demographic. Collectively, our research not only underscores the potential of AR in enhancing the digital journey of older adults but also offers \revision{implications} for its future application in the realm of human-computer interaction.
In sum, we make the following contributions:
\begin{itemize}
    \item Our study introduces a novel approach that employs AR as a supporting tool to assist older adults in exploring and learning smartphone applications. By showcasing the potential applications of AR in this context, we provide a direction for the design of AR-based smartphone learning aids tailored for older users.
    \item  Our exploration sheds light on the specific AR interaction modalities that resonate deeply with older adults. This provides an actionable design roadmap for future AR interaction design, aiming to enhance the inclusivity and user-centricity of AR experiences for the senior demographic.
    \item Recognizing and addressing the concerns and apprehensions of older adults regarding AR, we combined these findings with the UTAUT (Unified Theory of Acceptance and Use of Technology) and UTAUT2 framework. This fusion resulted in a comprehensive vision for the future of AR tools, ensuring their alignment with the needs, preferences, and concerns of older users.
\end{itemize}

\section{Related Work}
\subsection{Older Adults Learning Smartphone Challenges and Preferences}

\label{Preference}
As societies advance, the evolving digital landscape presents challenges to older adults, accentuated by natural declines in both physical and cognitive abilities. Vision impairments, common in the elderly, can make recognizing text on small screens challenging~\cite{czaja2006factors, Jin22I}. Furthermore, the broader click range seen in older adults can lead to operational errors, especially on crowded screens~\cite{jinsynapse2022,linwhy2018}. As a remedy, many show a preference for larger screens~\cite{pandyataptag2018, Pang2021Technology}.

Aging-related cognitive decline, such as forgetfulness or reduced concentration, can further hinder their digital learning process~\cite{Jin22I, guoolder2017}. Beyond this, older adults often grapple with foundational digital concepts—being unfamiliar with terms like ``digital photo album'' or ``cloud storage''-which may seem rudimentary to digital natives~\cite{behwhere2015, atkinsonbreaking2016}. Despite their eagerness to understand, existing resources frequently lack the clarity and detail this demographic needs~\cite{leunghow2012}. They prefer well-defined, step-by-step instructions and value interactive tutorials or in-app help videos~\cite{leunghow2012, Pang2021Technology}.

Interestingly, recent studies highlight older adults' shift towards more autonomous digital learning methods, like the trial-and-error approach~\cite{Pang2021Technology}. Yet, many remain apprehensive about potential device damage, a fear which often dictates their exploration behaviors~\cite{cyartoactive2016,atkinsonbreaking2016,mullerpractice-based2015}. Their explorations tend to be more focused, yet they display a heightened aversion to mistakes, especially those with perceived significant consequences~\cite{chinage2012,Jin22I,mullerpractice-based2015}. However, they may be more willing to try if they are supported to be more confident that errors would not lead to severe consequences~\cite{barnard2013learning}.

\subsection{Current Approaches to Supporting Older Adults Learning}
Within the domain of assisting older adults in learning digital technologies, research has gravitated around two primary strategies: social help learning and independent learning. Traditional in-person social assistance, faced with accessibility challenges, has been complemented by remote methods, including video chats and apps such as Meerkat, which offer articulate assistance mechanisms~\cite{mendelmeerkat2022}. The ``situated scaffolding method'' introduced by Cerna et al. champions a learner-focused approach, offering support in sync with the user's immediate context and objectives~\cite{cernasituated2022}. \revision{Although social learning is an important method, Sec~\ref{Preference} shows that older adults nowadays prefer more independent learning methods for flexiblility and confidence~\cite{Pang2021Technology, Jin22I}. Our work, therefore, focuses on aiding their exploration of apps independently, recognizing their willingness to try new approaches with better support.}

In the landscape of independent learning, research efforts have predominantly encompassed three spheres: instructional materials, device-embedded help features, and the trial-and-error methodology. Instructional materials have been studied extensively. Comprehensive guidelines, articulated with lucid language and supplemented by visual aids such as screenshots, have been of particular interest. Technologies like Live View, which offers real-time UI illumination, and demonstrative videos serve as valuable assets in this realm~\cite{behwhere2015,cyartoactive2016,leunghow2012}. Concerning device-oriented assistance, scholarly efforts have underscored the importance of personalized recommendations. The inclusion of relevant links, derived from a user's browsing history, ensures a guided experience, especially for older users who might have a limited technological backdrop~\cite{chinage2012}. Innovative tools, such as the tactile button presented by Conte et al. and TapTag by Pandya et al., facilitate guided exploration, making them invaluable in supporting independent learning endeavors~\cite{contehelp2019,pandyataptag2018}. In the realm of trial-and-error, the emphasis has shifted to creating a supportive environment for older adults. Adopting designs that echo real-world analogs aids in minimizing cognitive strain, and crafting simulations rooted in daily scenarios bolsters cognitive engagement~\cite{hollinworthrelating2010, palumbomicogito2021}. Recognizing the trepidation older adults might face, provisions such as exploratory modes have been proposed, enabling users to experiment while, safeguarding against irreversible changes~\cite{leunghow2012}. A progression in this direction is the work of Jin et al., where Synapse was introduced. This platform not only facilitates demonstrations by experienced users but also fosters a safeguarded space for novices to explore, learn, and rectify errors, accentuating the essence of exploration in the learning process~\cite{jinsynapse2022}.

Given this background, Augmented Reality (AR) emerges as a pivotal tool. Its interactive, immersive, and context-sensitive attributes can significantly enhance the exploration-based learning processes for older adults. Grounded in the comprehensive research on older adults' digital technology adoption, our study seeks to optimize independence in learning through AR-supported experiences.

\begin{figure*}[]
\includegraphics[width=5.7 in]{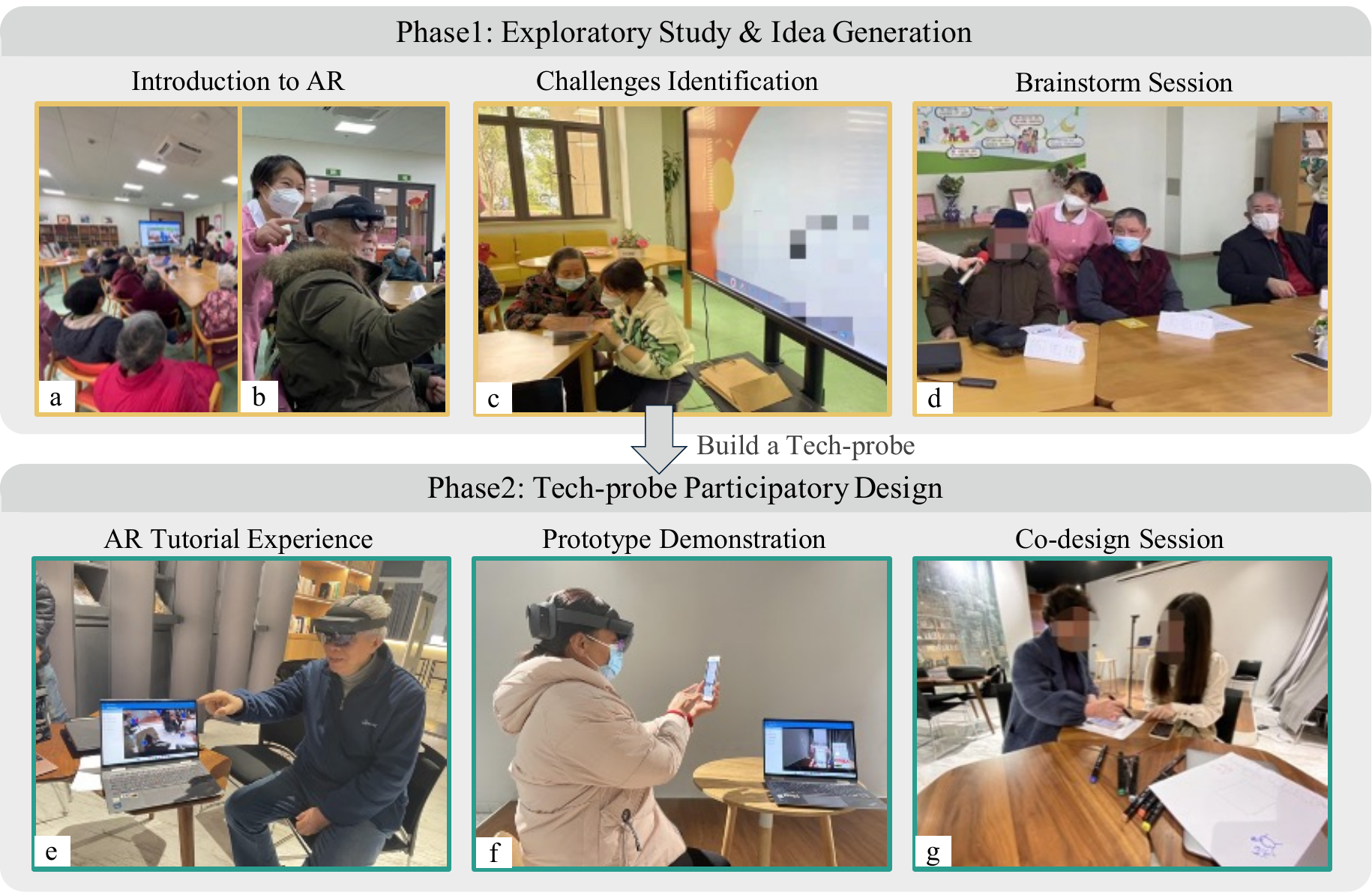}
\vspace{-0.2cm}
\caption{The Overview of the Two-phase Study. In Phase 1, (a) shows the introduction of AR's capabilities to participants, (b) shows their experience using AR, (c) shows the discussion of smartphone app exploration challenges, and (d) encapsulates their collaborative ideation on using AR to tackle these obstacles. Based on the findings from Phase 1, we built a tech-probe for Phase 2 participatory design. (e) and (f) display participants engaging with our AR tutorial and probe respectively, while (g) captures the essence of our participatory design approach with a participant sketching her ideas.} 
\Description{The figure has seven images in two rows. The upper row shows phase 1, the lower row shows phase 2. Phase 1 includes (a),(b),(c),(d), with (a) showing 18 participants sitting in the room, listening to the introduction of the researcher, (b) showing one participant experiencing AR, (d) showing one participant speaking to the microphone to share his idea with others. Phase 2 includes (e),(f),(g), with (e) showing one participant wearing AR to interact in mid-air, (f) showing one participant handling her phone with her left hand, and experiencing our probe, (g) showing the participant is talking with the researcher and sketching her idea to co-design the future AR tool.}
\label{fig: Overview}
\end{figure*}

\subsection{Potential of AR in Supporting Older Adults Learning}

Emerging from the confluence of display techniques and computer vision methodologies, such as SLAM, hand tracking, and object detection, contemporary head-mounted AR devices seamlessly blend the virtual with the tangible. This synergy not only amplifies the visual dimension but also redefines the boundaries of interaction, thereby reimagining the paradigms of learning.

\textbf{AR's Transformative Impact on Interactive Surfaces.} AR technology defies the limitations of traditional interactive platforms. The tangible realm is no longer a constraint, as demonstrated by Suzuki's work, which extrapolates interaction beyond the physical realm, introducing dynamic, context-aware, and enhanced virtual interfaces overlaying physical screens~\cite{suzuki2022augmented}. Innovations like those of Reipschläger et al. exemplify this shift, wherein virtual content is organically integrated into our environment, thereby fostering intuitive interactions~\cite{reipschlager2020augmented,reipschlager2019designar}. The potential of this expanded interaction canvas is further echoed in research focusing on CAVE systems and smartphones~\cite{cruz1992cave, nishimoto2019extending, hartmann2020extend}. Such expansions could particularly benefit older adults, who might find the compact and intricate interface designs of contemporary smartphones challenging~\cite{kane2008slide}. By stretching the interaction frontier, AR ushers in the era of custom-tailored and accessible control architectures, thereby catering to the distinct needs and potential limitations of older individuals.

\textbf{Redefining Learning with Visual Cues.} With its potential to superimpose visual hints, AR rejuvenates tutorial methodologies, making learning not just more insightful but also more immersive and delightful. Findings by Klopfer and Squire underscore this, revealing how AR's capacity to provide situational, context-rich information can simplify intricate concepts and catalyze collaboration~\cite{klopfer2008augmented}. The superiority of AR-driven learning over conventional textbook-based methodologies is empirically validated~\cite{albrecht2013effects}. Its efficacy is further reinforced in specialized domains such as medical education, where AR significantly enhanced anatomical comprehension and retention~\cite{kucuk2016educational}. Initiatives like the InstruMentAR system by Liu et al., AdapTutAR by Huang et al., and RobotAR toolkit by Villanueva et al. are a testament to AR's transformative potential~\cite{liu2023instrumentar, huang2021adaptutar, villanueva2021robotar}. However, a conspicuous void exists in the domain of smartphone application exploration for older adults using AR, a research gap that our study endeavors to bridge.

While the promise of AR in revolutionizing the tech-learning curve for the older generation is unmistakable, a detailed understanding of their aspirations and preferences for AR remains elusive. This study adopts a probe-based participatory design methodology to illuminate AR's potential in optimizing older adults' experience of technology learning.

\section{Method}
To address our RQs, we conducted a two-part study: (1) a workshop on investigating what challenges older adults face most when learning smartphones and where AR can be applied to offer support, (2) technology probe-based participatory design sessions to brainstorm and develop an innovative AR support tool collaboratively. The insights gleaned from the first part study directly informed and shaped the design of the probe in our subsequent second part study. A summary of the studies is presented in Figure~\ref{fig: Overview}. This research received approval from the university ethics review board. 

\subsection{Data Analysis}
We recorded the workshop and design sessions and aimed to identify themes related to older adults' experiences, perceptions toward the AR experience, and how they imagine incorporating AR in the learning smartphone process. The video recordings were transcribed through an online transcribing tool and then manually reviewed by the authors to correct for errors. The transcripts were coded using an open-coding approach~\cite{corbin2014basics}. \revision{For example, one significant code was ``leveraging AR to mitigate memory declines of multi-app exploration'', defined as the design where participants conceived using AR's exteneded surfaces to support their exploring process with multiple apps. Two examples are as follows: ``I often forget which app should be used for the next step and AR may help remind me the next app for my task'', ``I like to compare items between several apps, but I can not remember all of them, and had to check back and forth.''}. Two of the authors reviewed the initial transcripts \revision{independently}, and then developed and applied codes in an iterative process until reaching an agreement. \revision{This iterative process ensured a consistent and comprehensive coding scheme.} The codes were subsequently organized into clusters that represented the emerging themes from the study data. \revision{For example, ``leveraging AR to mitigate memory declines during multi-app exploration'' combined with similar codes such as ``leveraging AR to help focus on the current exploration for those `attention declines'' formed the theme ``Leveraging AR to Reduce Cognitive Burden of Older Adults during Smartphone Apps Exploration and Learning.''} Once all potential themes were thoroughly discussed and reviewed, the final themes were selected as the study's findings.


\section{Phase 1: Exploratory Study \& Idea Generation}

\subsection{Study Design}

In phase 1, we aimed to explore the critical challenges participants encountered when learning smartphone apps by exploring and brainstorming how AR could be used to tackle these challenges. It was conducted in a nursing home with 18 older adult participants who lived in the nursing home or nearby local communities. The study lasted approximately 90 minutes and included three parts. 

\textbf{Part 1, Introduction to AR}: We introduced older adults to AR technology by showing them a prepared introduction video of how AR works and letting them try an AR task of picking up a flower and putting it into a vase through AR Hololens 2\footnote{https://www.microsoft.com/zh-tw/hololens/hardware}, which includes basic AR interactions like selecting and moving. This helped them understand the potential of AR and how it could be integrated into their daily lives.

\textbf{Part 2, Identifying Challenges}: We inquired about the difficulties they faced while exploring smartphone apps. To help them better recall the challenges, we gave them a health management app with rich features as an example, then asked them to try using it and point out where they experienced difficulties. A health management app was chosen because older adults are more likely to use technology that is integrated into their daily lives, such as health monitoring devices~\cite{mitzner2010older}.

\textbf{Part 3, Brainstorming Session}: We held a brainstorming session with the participants to encourage them to think about how AR could be used to overcome the challenges they encountered while using smartphones. We listed the challenges they raised in Part 2, and for each challenge, we allowed participants to contribute ideas on how AR can be used to tackle these challenges. 

\subsection{Participants}
We recruited eighteen participants (10 males, 8 females) aged 65-94 (M = 77.1, SD = 8.5). All participants had used smartphones before and reported that they encountered challenges when learning smartphone apps from the preliminary questionnaire used for collecting their demographic information. None of them used AR applications before but all were open to new technologies.

\subsection{Findings from Phase 1: Smartphone Challenges During Exploration and Possible AR solutions}
Our participants showed extreme interest in AR and actively experienced wearing the AR device. We identified four core challenges for smartphone app exploration: 

\textbf{\nnboxx{C1}: Limited size of text on the screen.}
Participants reported that their visual acuity was reduced, causing them to spend more effort and time trying to read the small text on the screen. To tackle this problem, participants mentioned having a larger panel with bigger font sizes in AR floating beside the touchscreen.

\begin{figure*}[]
\includegraphics[width=5.5 in]{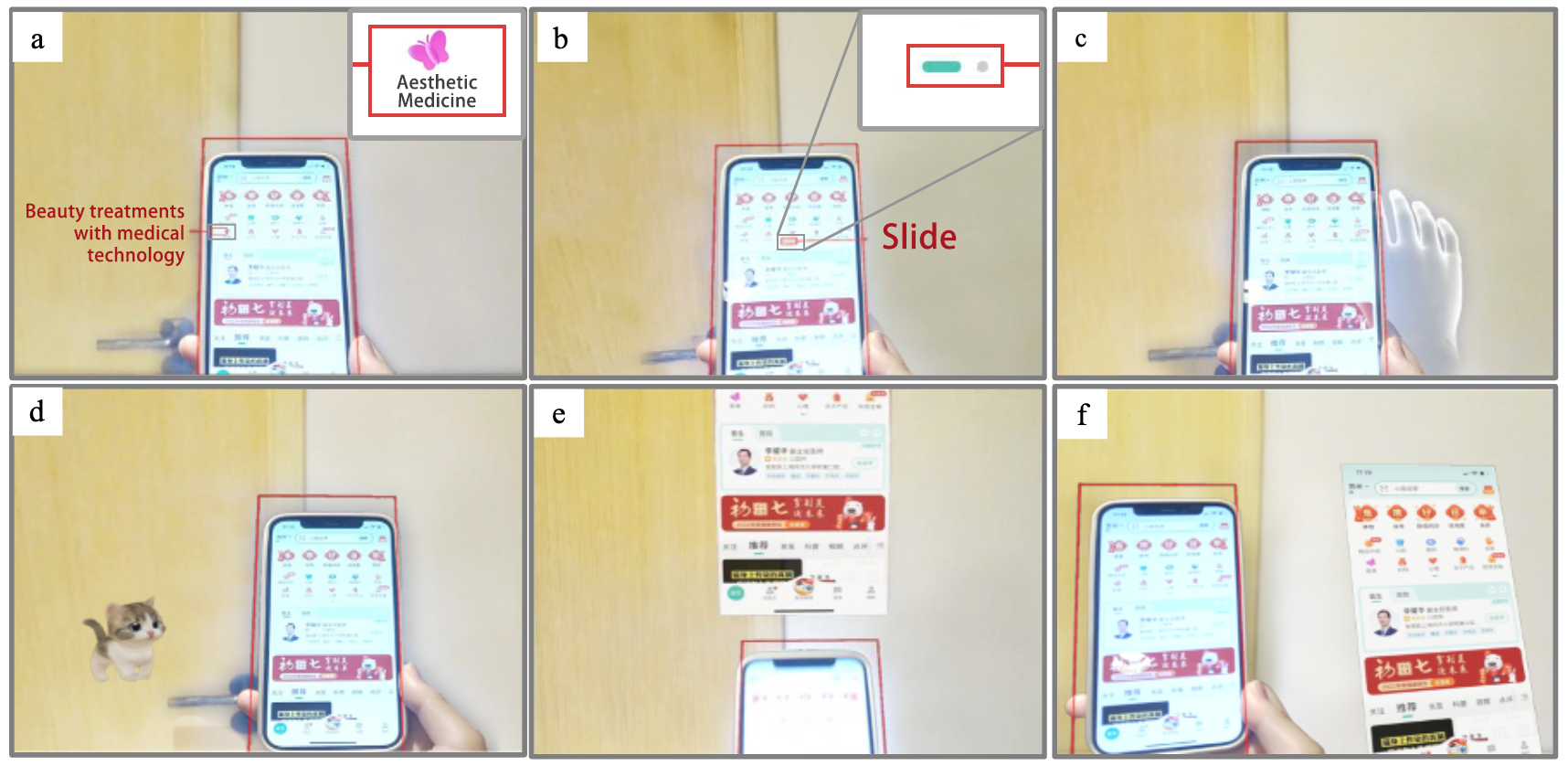}
\vspace{-0.2cm}
\caption{Features of the Design Probe. (a) enlarges and explains a terminology (in this example, it is ``aesthetic medicine'') and shows the enlarged explanation on the side of the smartphone. (b) highlights the hidden feature and explains it (in this example, it means ``sliding'' ) on the side of the smartphone. (c) uses a ghost hand animation to demonstrate the feature's meaning. (d) shows a 3D cat as an accompanying assistant. (e) demonstrates a vertical extended screen on the top of the smartphone. (f) demonstrates a horizontal extended screen on the side of the smartphone.} 
\Description{The figure contains six images distributed in two rows, where each image shows the screen of a smartphone app and corresponds to a feature in the design probe. (a) enlarges and explains the term aesthetic medicine. (b) highlights the hidden feature and explains it with the text ``sliding'' beside the screen. (c) shows a ghost hand animation to demonstrate the meaning of sliding. (d) shows a 3D cat as an accompanying assistant. (e) shows a vertical extended screen. (f) shows a horizontal extended screen.}
\label{fig: Features}
\end{figure*}

\textbf{\nnboxy{C2}: Low information scent of features.}
Participants also found it hard to figure out how to find certain functions, which is due to the low information scent of features. To better accommodate more features or functions in one app, the current design of apps uses a ``hidden'' design, which requires the user to trigger certain functions with gestures like sliding. Often, ``hidden'' design features are subtle, and characterized by minimalistic symbols, as shown in Figure ~\ref{fig: Features} b the sliding feature. For participants unfamiliar with extensive smartphone app usage, it becomes challenging to discern their interactivity, making them overlook the additional functions nestled behind such features. Participants reported that they wanted such hidden features to be highlighted and the operations that are possible to be explained. 

\textbf{\nnboxz{C3}: Difficulty understanding terminologies.}
Participants reported that sometimes they were unsure what would happen if they clicked certain icons. Due to the expanding functions in many apps, many used shortened terms to reduce screen space, making it even harder to understand for those who were already not familiar with the app. One participant mentioned asking others for help, \textit{``So in the end, I only use the one [function] I can understand, if I had to use others, I will ask [family members] to describe it to me, but I cannot ask them too often, because they are busy as well.''} In severe cases, participants mentioned they just gave up using the app. To address this issue, participants felt that having a more detailed explanation of the terminology in plain language floating beside the screen would be helpful. 

\textbf{\nnboxo{C4}: Low confidence when exploring independently.}
Participants reported their anxieties when exploring smartphone apps independently. They felt less confident because they were experiencing a cognitive decline that they perceived in their daily lives. For example, a participant mentioned, \textit{``my brain is not as sharp as before, and I often forget things in recent years, I am afraid that I might make mistakes.''} They expressed extra caution when conducting operations related to payment, which could have serious consequences if a mistake was made. Participants mentioned that a second virtual screen that displays the result of an operation could act as a trial run. If the virtual screen matches their expectations, they would feel more relieved to operate on the real one. Another approach they suggested was having an accompanying assistant who could answer their questions when they felt unsure.

\section{Phase 2: Tech-probe Co-Design Workshop}

\subsection{Design Probe Implementation}

Based on the challenges uncovered (C1-C4) from Phase 1, we developed a probe of an AR application designed to support older adults in their exploring process with the following features:  

\begin{enumerate}
    \item Enlarge the font size in two ways, enlarge certain text, and enlarge the whole screen to address \revision{\nnboxx{C1}: limited size} (Figure~\ref{fig: Features} a, e, f). 
    \item Explain the terminology to address \revision{\nnboxz{C3}: difficulty understanding} (Figure~\ref{fig: Features} a). We added a text explanation of the icon beside the smartphone. We chose a red color for the demonstration because it shows more clearly in AR under our laboratory settings. 
    \item Highlight hidden feature and add illustration to address \revision{\nnboxy{C2}: unclear features} (Figure~\ref{fig: Features} a, b). Although our participants only mentioned text explanations for highlighting the hidden feature, we added an animation of a ``ghost hand'',  which was found to be older adults' most preferred visual prompt~\cite{Williams2021Augmented}.
    \item Add extended screen to address \revision{\nnboxo{C4}: low confidence} (Figure~\ref{fig: Features} e, f). We provided two different locations for the virtual screen, above and to the right of the physical smartphone. 
    \item Add an accompanying assistant to address \revision{\nnboxo{C4}: low confidence} (Figure~\ref{fig: Features} d). As most of our participants reported that they preferred accompanying animals' avatars, we placed a cute 3D cat as a demo.
\end{enumerate}


We implemented the AR probe in Unity and deployed it using Microsoft HoloLens 2. To display AR content in reality, we selected a standard position for holding a mobile phone. We utilized Mixed Reality Toolkit\footnote{\url{https://learn.microsoft.com/en-us/windows/mixed-reality/mrtk-unity/mrtk2/?view=mrtkunity-2022-05}} to synchronize the AR world coordination system with reality and maintain the virtual content's consistency in space. The ``ghost hand'' animation was constructed using MRTK's default hand model.
To allow changing of the features of the AR probe in real-time during the design probe, we built a configuration server, which runs on a laptop computer, using Unity Netcode\footnote{\url{https://docs-multiplayer.unity3d.com/netcode/current/about/}}, to communicate with the HoloLens 2 AR headset. In this way, we can control which feature to show up in the design probe through our built server on the laptop computer.


\begin{figure*}[]
\includegraphics[width=4.5 in]{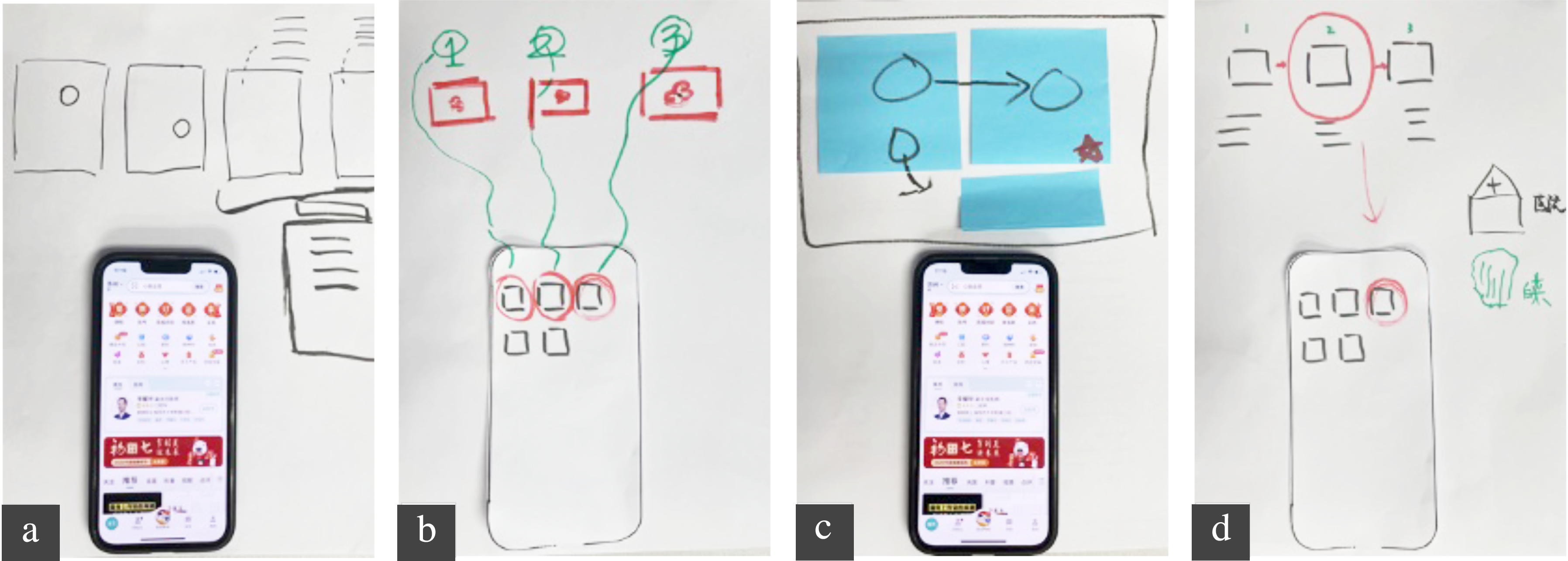}
\vspace{-0.2cm}
\caption{Older Adults' Drawing Samples. (a) shows that they would like AR tools to display their operation flow and summarize previously filled-in information so that in the confirmation stage of a task, they can check the information more easily and confidently. (b) shows that they wanted AR to present three apps on the same screen for them to compare the prices of vegetables. (c) shows that they wanted AR to record and save their trial-and-error process so they could return to any saved state. (d) shows that they wanted a custom icon to record multi-app usage and show which states they were currently in.} 
\Description{(a) shows four rectangles on top of the smartphone, representing screenshots of their process, and below those, there is another rectangle recording all filled-in information. (b) shows three rectangles, representing three apps on top of the smartphone, and showing the vegetable's prices for comparison. (c) uses sticky notes on top of the smartphone, each sticky note represents a stage of trial-and-error, and for the nearest recovery stage, there is a red star. (d) shows custom icons like hospital icons, and vegetable icons, representing their daily activities. When choosing one icon, it shows the whole process of the task, and organizes them by independent app, meanwhile, it shows a red circle of the current app and will give a hint on the smartphone as well.}
\label{fig: multi}
\vspace{-1.5em}
\end{figure*}

\subsection{Study Design}

We initiated our research with pilot studies involving 2 older adults, which helped us fine-tune our co-design process. The study was hosted at a community library, a locale that was both familiar and easily reachable for the participants.

We used the probe for participants to experience receiving assistance in AR and co-design with the researchers to tackle their challenges and better facilitate their exploring process. The study lasted approximately 90 minutes and included three parts. 

\textbf{Part 1, Preparation and AR Tutorial Experience}: Considering that our participants did not have experience with AR, we gave a tutorial on AR to familiarize them with the basic interactions~\cite{microsoft2019hololens}. We also observed their interaction process during the tutorial and asked how they felt about the interaction. This not only allowed participants to better understand what could they do in AR but also allowed researchers to understand older adults' perceptions of AR interactions. \revision{To ensure that participants understand and become familiar with the AR interactions, we first present all interactions as depicted in Figure ~\ref{fig:interaction}. Subsequently, we ask them to describe these interactions both verbally and through gestures, as a means of verifying their comprehension.}

\textbf{Part 2, Tech-probe Experience and Feedback Gathering}: We then let older adults put on the Hololens 2 to experience how AR supports smartphone app exploration. We let them try using the same app as Phase 1, and when they encountered challenges, we presented our relative AR features to them and explained how it may help. For the rest of the features, which participants did not experience, we presented them at the end. For each feature, we asked them about their experiences, what they liked, what they did not, and what improvements they thought could be made. 

\textbf{Part 3, Co-Design Session}: Finally, we held a co-design session with the participants. We prompted participants by recalling their encountered challenges, and our design features of tech-probes, to suggest changes and add new features to support their learning process of smartphones. We also encouraged them to draw their ideas on paper to better express themselves.

In the practical execution of Phase 2, individual participatory design sessions were held, with each of the two researchers working separately with one participant. Following the individual sessions, we facilitated a group discussion, which involved both researchers and the two participants, to further refine the co-design solutions. \revision{In this process, each participant expressed their ideas first while the others gave comments and suggestions. We discussed the ideas until everyone was satisfied with the design.} We initially planned for 16 participants across 8 sessions, but unfortunately, one participant did not attend due to unforeseen personal circumstances.

\subsection{Participants}
Fifteen participants (6 males, 9 females) aged 65-94 (M = 75.7, SD = 9.5) were recruited in Phase 2. Seven were volunteers from Phase 1, while others were recruited through advertisements on social media. All had diverse backgrounds and education levels. Based on the definition of levels of smartphone usage expertise~\cite{leunghow2012}, there were one beginner, four novice users, seven intermediate users, and three advanced users. Recruiting criteria were: (a) Participants had the ability to use smartphones and encountered challenges when learning smartphone apps by exploration. (b) Participants should be willing to experience AR and explore how AR could be applied to support the smartphone learning process. 

\section{Findings of Tech-probe Co-Design Workshop}
We first present the key findings of how AR can be applied to support participants in learning smartphone apps to answer the first research question. We then present the participants' preference for interactions with AR to answer the second research question. Figure ~\ref{fig: multi} contains selected sketches from participants. Since we also have those who responded verbally instead of sketching, we created illustrations to make the point clearer.


\subsection{Possibilities of AR to support older adults learning smartphone apps}

We identified four general directions in which AR could be leveraged to support older adults in smartphone learning, which are reducing their physical burdens, lowering their cognitive burdens, facilitating trial-and-error, and bridging the knowledge gap.

\subsubsection{Leverage AR to reduce older adults' physical burden of learning smartphone apps}

We reported how AR can support tackling two general challenges in older adults' physical aspects of learning smartphone apps.

\textbf{AR can provide user-friendly operations for reduced fine motor skills and tactile sensitivity}.
Fine motor skills are necessary in accurately executing touch gestures on smartphone screens. When exposed to our AR probe, participants commended the benefits of diminished operational demands, notably via the extended screens. The scalability of this extended screen, allowing touch targets to be magnified, emerged as particularly beneficial. P3 remarked, \textit{``I often click on the wrong icon since they are too close to each other, but I will not click wrong if they are magnified at this scale.''} Additionally, the issues with unreliable touch responses on smartphones were underscored by P6, who stated, \textit{``I often find myself tapping repeatedly to ensure I clicked the icon successfully. Sometimes it failed to respond, and once, I clicked too much, my phone even shut down. If I can use it through the helmet [AR Hololens] instead of smartphones, I will choose this.''} Moreover, participants mentioned that performing gestures to switch between apps on small smartphones was physically demanding. Participants proposed using AR to display recently used apps beyond the smartphone interface, which would streamline the app-switching process via the interaction with floating application icons. However, the transition to AR wasn't without its challenges for the older demographic. A recurrent point of contention was the intangibility of mid-air interactions, which we report in detail in Section ~\ref{sec:interaction-findings}.

\begin{figure*}[]
\includegraphics[width=5.7 in]{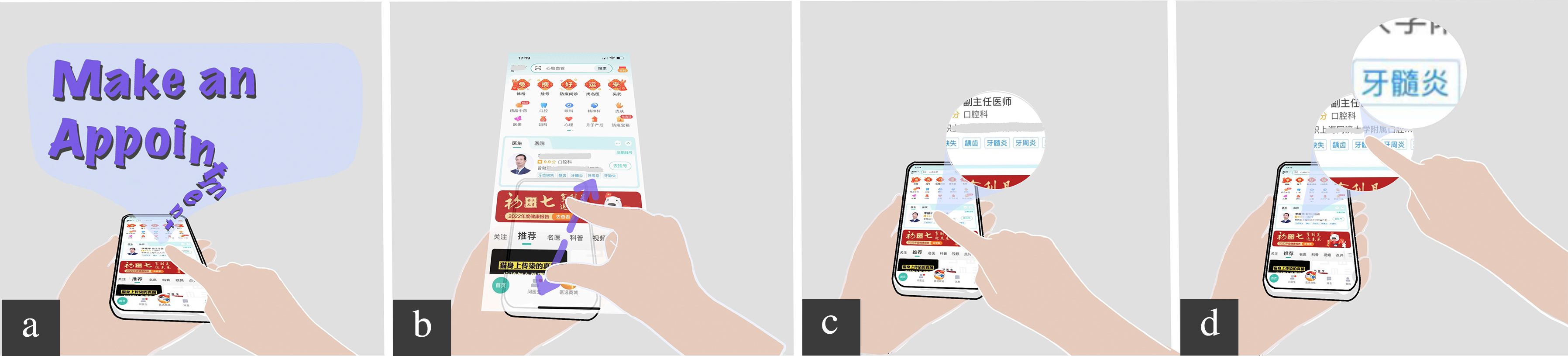}
\vspace{-0.2cm}
\caption{Participants' Vision of Employing AR to Augment Visual Interactions. (a) illustrates 3D words in an exaggerated brush style that seem to ``jump out'' from the display. (b) depicts a gesture-induced screen magnification. (c) visualizes a finger simulating a magnifying effect. (d) reveals a multi-tiered magnification process.}
\Description{In (a), the phrase "Make an appointment" appears with each individual letter seemingly leaping off the screen. (b) captures an AR-enhanced gesture that enlarges the entire display. (c) represents a hand, with the specific point it targets magnified, emulating the effect of a physical magnifier. (d) builds upon (c) by further magnifying a specific word within the previously enlarged section from (c).}
\label{fig:Visual}
\end{figure*}

\textbf{AR can enhance visual experiences for those with decreased visual acuity}.
The constraints of smartphone screens, particularly on smaller devices, present legibility challenges for individuals with declined visual acuity. Participants echoed these challenges, revealing a common solution: increasing font size. This approach, however, occasionally resulted in distorted layouts, extending icons over multiple screens, and complicating user interaction. In contrast, AR offered the promise of enhancing content clarity, as showcased in our demo.

For small content such as text underneath an icon, we projected them beside the screen, receiving unanimous approval from participants. They felt this enhanced visibility and suggested the inclusion of accompanying audio for a richer experience. Participants also gave suggestions on content representation: P3 wanted content that ``jumped out'' for greater engagement, while P5 desired a brush-styled font, as Figure~\ref{fig:Visual} a shows. Interestingly, P10 expressed that she was used to having a magnifier, so she hoped that her finger could be a magnifier that automatically magnified the words beneath, as Figure~\ref{fig:Visual} c shows. Bolstered by our extended screen concept, P2 proposed magnifying the entire screen, thereby proportionally increasing the text size, as Figure~\ref{fig:Visual} b shows. P7, despite having good vision, expressed a preference for larger text to safeguard future eyesight. The diverse feedback culminated in P4's integrated solution: enlarging the entire screen for recreational apps like games, while selectively magnifying crucial parts for functional apps like health trackers. P12 introduced another layer of customization, envisioning enlarged UI sections based on user preference, enabling further iterative magnification as Figure~\ref{fig:Visual} d shows.

Addressing inconspicuous feature cues, participants favored an on-demand display over automatic visibility, which means that they would like to manually trigger the feature. P3 valued post-exploration feature identification, while P11 highlighted the distractions of spontaneous notifications. In terms of visual emphasis, participants found the demo's color-highlighting approach obscured feature details, suggesting a more pronounced highlighting format. Notably, they exhibited a dual attitude towards AR visualization: while unexplored features should retain their original appearance on the smartphone screen, selected ones could harness AR's expansive screen capacity for enriched detail presentation.

\subsubsection{Leverage AR to reduce older adults' cognitive burden of learning smartphones}

\textbf{AR can facilitate smartphone app exploration for those with decreased working memory capacity.} 
Participants recounted difficulties in executing intricate tasks on their smartphones, particularly when these tasks spanned multiple apps. To combat forgetfulness in these extended, multi-app endeavors, participants, like P15, proposed AR-aided lists for marking off completed steps---a digital parallel to traditional note-taking.

\begin{figure*}[]
\includegraphics[width=5.7 in]{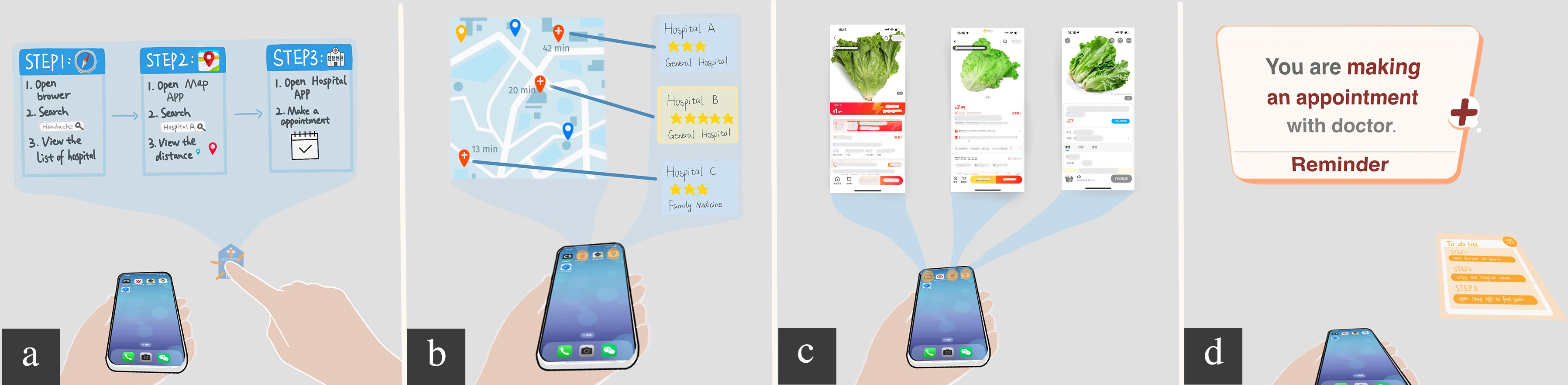}
\vspace{-0.2cm}
\caption{Participants' Vision of Utilizing AR to Alleviate Cognitive Load. (a) illustrates a  custom icon for multi-app tasks, organized by individual apps. (b) demonstrates the integration of diverse information from multiple apps onto a single display. (c) portrays the simultaneous presentation of three apps, designed for efficient price comparison. (d) highlights a reminder board that aids in refocusing the user's attention to their current task following interruptions.}
\Description{In (a), a hand interacts with a hospital icon, leading to the display of three separate screens, each detailing the steps and objectives of its corresponding app. (b) showcases a map interface merged with hospital details from a search engine, presented on a unified screen. (c) displays three parallel screens, each presenting vegetable prices, facilitating effortless comparison. (d) features a prominent board with the message "You are making an appointment with a doctor. Reminder." Adjacent to it, there's an editable to-do list, allowing older users to modify the reminder board's content.}
\label{fig:cognitive}
\end{figure*}

A recurring challenge was the cognitive load of multi-app usage. P8's medical journey exemplified this: starting with symptom research on a search engine, shifting to a map app to identify nearby hospitals, and finally accessing the chosen hospital's dedicated app for appointment scheduling. The need for multiple switching and information recall from different apps led to fatigue and forgetfulness. Inspired by our AR probe's extended screen, P8 proposed custom icons for regular app sequences, as Figure~\ref{fig:cognitive} a shows. Clicking a bespoke hospital icon, for example, would display an entire operational flowchart categorized by different apps. This could streamline the task processes, preventing feelings of being overwhelmed or missing out on vital information such as hospital operating hours. Moreover, when coming to the next app, it can also help older adults select the right one from similar app icons. 

During the process, another significant challenge for older adults was comparing data across multiple apps to make informed decisions. For instance, when choosing a hospital, factors like reputation and location are crucial. However, gathering this information often required toggling between different apps, demanding a considerable cognitive load for memorization and decision-making. P7 highlighted a potential solution: an AR-enabled feature that consolidates relevant data from various apps onto a single extended screen, as Figure~\ref{fig:cognitive} b shows. This would facilitate direct comparisons, alleviating the burden of recalling and cross-referencing information from disparate sources. The dilemma of juxtaposing information was not limited to medical choices. P10 described the frustration experienced when comparing vegetable prices across multiple shopping apps. After frequently losing track of initial price checks, the repetitive process sometimes led her to abandon the task. To address this, participants suggested employing AR to simultaneously present all app interfaces, spotlighting pertinent details like pricing, as Figure~\ref{fig:cognitive} c shows. As P10 noted, such an approach is reminiscent of ``having multiple windows open on a computer,'' simplifying decision-making through concurrent visibility.

\textbf{AR can enhance focus for older adults who experience a decline in attentional resources}
Participants frequently expressed frustration with interruptions during their app exploration, citing disturbances like pop-up ads, app notifications, and incoming calls. Drawing inspiration from our AR demonstration, which showcased extended screens and floating text, many participants proposed a continual display of their current task and chosen app, helping them stay focused. P7 suggested, for instance, that a prominent reminder, such as \textit{``I'm making a doctor's appointment''}, could be invaluable. Echoing this sentiment, P8 emphasized the advantage of placing these reminders at eye level, as Figure~\ref{fig:cognitive} d shows. Given that many participants naturally glanced upwards during moments of recollection, this positioning could facilitate seamless task resumption. An unexpected benefit noted by some participants was that looking upwards might serve as a good neck exercise. Additionally, the challenges weren't solely related to external interruptions. Due to diminishing attention spans, participants often faced confusion when selecting between similar apps. P14's experiences reflected this, as she frequently opened incorrect apps, leading to frustration. She believed that AR could act as an assistive tool, guiding users by highlighting the appropriate app for their intended task.

\subsubsection{Leverage AR to support older adults' trial-and-error and mitigate mistake anxiety.}

\textbf{AR can establish a digital sandbox for older adults to explore app functions safely}
Our demo presented an extended screen and allowed older adults to conduct trials on the virtual surface. Participants felt that it made them more confident to conduct operations on the real phone if they had tried it before (e.g., \textit{``It allows me to familiarize myself with these functionalities without the fear of making mistakes.''}-P3).

\begin{figure*}[]
\includegraphics[width=5.7 in]{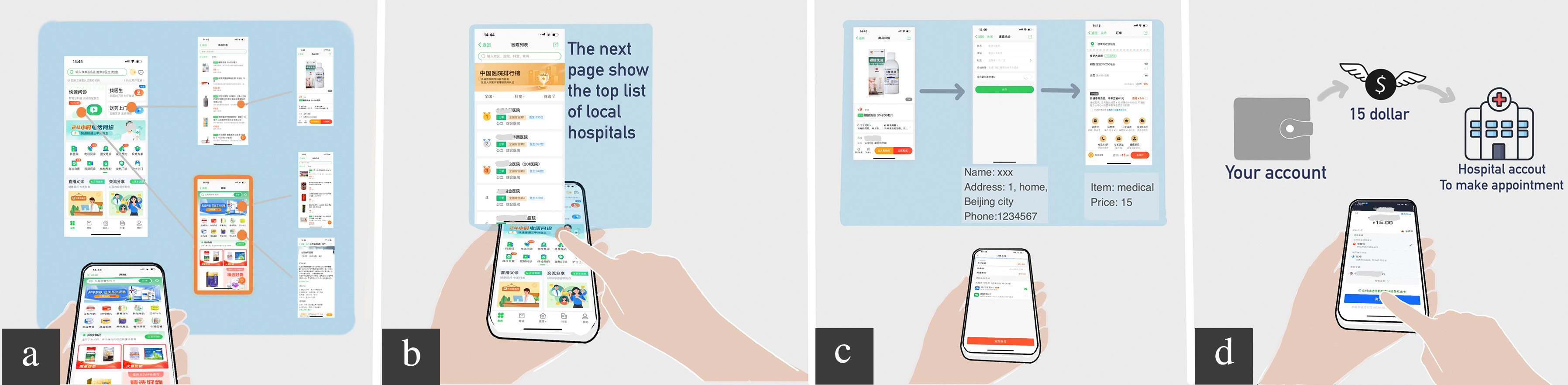}
\vspace{-0.2cm}
\caption{Participants' Insights on Utilizing AR for Supporting Trial-and-Error Processes. (a) illustrates the tracking of users' trial-and-error routes, capturing principal actions and saving intermediate states for easy revisiting. (b) displays a live preview of the subsequent page when an icon is selected. (c) represents a scenario where a task, especially one requiring confirmation, lays out each step and lists all entered key details for verification. (d) highlights a real-world scenario to boost comprehension and confidence during payment confirmation.}
\Description{In (a), a series of screenshots represent various stages of operation with the active stage accentuated by a red rectangle. (b) portrays a hand tapping an icon, subsequently unveiling a sneak peek of the next page. (c) presents three snapshots symbolizing the progression of operations, underlined by the associated entered data. (d) showcases a purse, accompanied by user-specific bank details, the precise transfer amount, and an icon of a bank denoting the recipient, providing a tangible context to the digital transaction.}
\label{fig:trial-and-error}
\end{figure*}

\textbf{AR can present and save different states \revision{during} explorations and enable states to be recovered.}
Participants frequently faced moments of uncertainty while navigating apps, with deeper exploration sometimes leading to feelings of disorientation and the need to restart their tasks. \revision{We employ the term 'different states' to describe the varying phases of user progress. These states denote distinct points or stages within the user's exploration journey.} P5 highlighted the importance of saving one's progress at the beginning so that in moments of confusion, one could return to a known point rather than starting afresh. Building on this, P6 advocated for the ability to save multiple states or checkpoints within the AR system, facilitating a more flexible, trial-and-error approach without redundancy, as Figure~\ref{fig:trial-and-error} a shows. P12 emphasized the potential benefits of visual aids to help users track their navigational routes. By visualizing their paths, users could adopt a more methodical exploration strategy, avoiding repetitive actions and fostering more efficient learning. Reflecting on the emotional toll of such experiences, P12 shared, \textit{``Navigational errors often make me feel distressed. I often panic, and sometimes I try and try and try, sometimes repeating the same flow, until I lose confidence. I felt really awful that time, and do not want to do it again by myself.''}

\textbf{AR can mitigate older adults' navigational uncertainties by providing a preview of the next screen of the app.}
Participants suggested presenting a preview of the next page to build a reassuring environment and reduce their anxiety. As expressed by P8, \textit{``When I tap something, I always have a split second of panic wondering if I've done something irreversible.''} An AR preview acts as a safety net, allowing them to anticipate and validate their actions. P14 resonated, "Knowing what comes next, even just a glimpse, reduces the anxiety of navigating new apps or features." Besides previewing the pictorial page, P9 mentioned the utility of brief textual descriptions overlaid on significant parts of the preview, as Figure~\ref{fig:trial-and-error} b shows, \textit{``Just a word or two about what a button does would be immensely helpful.''} P12 further confirmed the anticipated benefits of enhancing learning and reducing errors, \textit{``Seeing a sneak peek of what's next helps me mentally prepare, making it easier to learn and remember. With a clear sense of direction, I may make fewer mistakes.''} 

\textbf{AR can enhance older adults' confidence in the confirmation stage.} 
The ``confirmation'' step in digital tasks can be particularly anxiety-inducing for older adults, as it often marks the point of no return, for example, when confirming the payment. P1 voiced concerns about the lack of summaries prior to finalizing actions, noting that the absence of a step-by-step review makes her retrace her steps as a precaution. P1 stated, \textit{``I feel afraid to click on the confirmation icon. Without a summary of every step I have taken and information I have filled in, I usually click cancel, and trace back to check each step, which is time-consuming and makes me feel more worried.''} She expressed a preference for a visual overview as shown in Figure~\ref{fig:trial-and-error} c, allowing for a quick and confident review. Meanwhile, P10 suggested highlighting critical or high-risk actions to ensure users are fully aware of the implications of their choices. P14 favored real-world analogies, likening digital payments to physically transferring money, an idea reflected in Figure~\ref{fig:trial-and-error} d, which makes the digital experience more tangible and relatable. P14 stated \textit{``I feel more sure about what I am doing. I feel easier to confirm my information in this way instead of reading a long text with terminologies.''}

\textbf{AR can help mitigate older adults' security concerns of exploring apps.} 
Older adults are wary of security vulnerabilities in mobile apps due to the increasing cyber threats reported in the news. They hope AR can offer visual warnings for risky operations. A notable idea from P12 is for the AR assistant to transform into a ``policeman'' during risky operations, indicating caution. The avatar of the AR assistant also evoked diverse reactions. While some participants preferred a more approachable, comforting presence like that of a household pet to foster a sense of trust and ease, they also expressed a desire for this avatar to dynamically adapt. Specifically, during potentially risky tasks, the avatar could metamorphose into a protective figure, like the aforementioned policeman, to underscore the gravity of the operation. Moreover, participants mentioned using animations or spatial visualizations to notify them of risky operations. For example, P14 said \textit{``I hope that it could help me get out of certain states, if I were scammed and about to transfer money, it can alert me and let me get out of the mood.''} Interestingly, many participants feel that AR, being separate from their phones, is inherently safer and less invasive, thus boosting their trust.

\subsubsection{Leverage AR to Bridge Knowledge Gap for Older Adults When Learning Smartphone Apps}

\textbf{AR can explain the complex understanding of the terminology and illuminate the ambiguity of interface symbols in diverse ways.}
In the rapidly evolving technological landscape, older adults often grapple with new terms. Our AR demonstration proposed clear and short textual explanations alongside unfamiliar terms. Participants favored concise text over lengthy descriptions to prevent memory lapses. While text was essential, participants also appreciated other mediums like animations and voice. The immersive animations and spatial effects of AR were not only informative but also entertaining for them. P3 was enamored with the idea of words coming to life, P7 imagined complete immersion in a ``smartphone world'', while P10 saw AR as a dynamic learning tool, superior to lessons from younger relatives.

Hidden app features, often symbolized by small icons, tend to be overlooked by older users. To address this, our demo showcased methods like highlighting these features with both text and animated cues. While text was favored for its clarity, the timing of these feature pop-ups was crucial to avoid distractions. Moreover, participants reflected that animations may not be a good choice because they needed more time to comprehend what the animation referred to. 

\subsection{Interaction preferences of older adults in AR for smartphone exploring}
\label{sec:interaction-findings}
\begin{figure*}[]
\includegraphics[width=5.7 in]{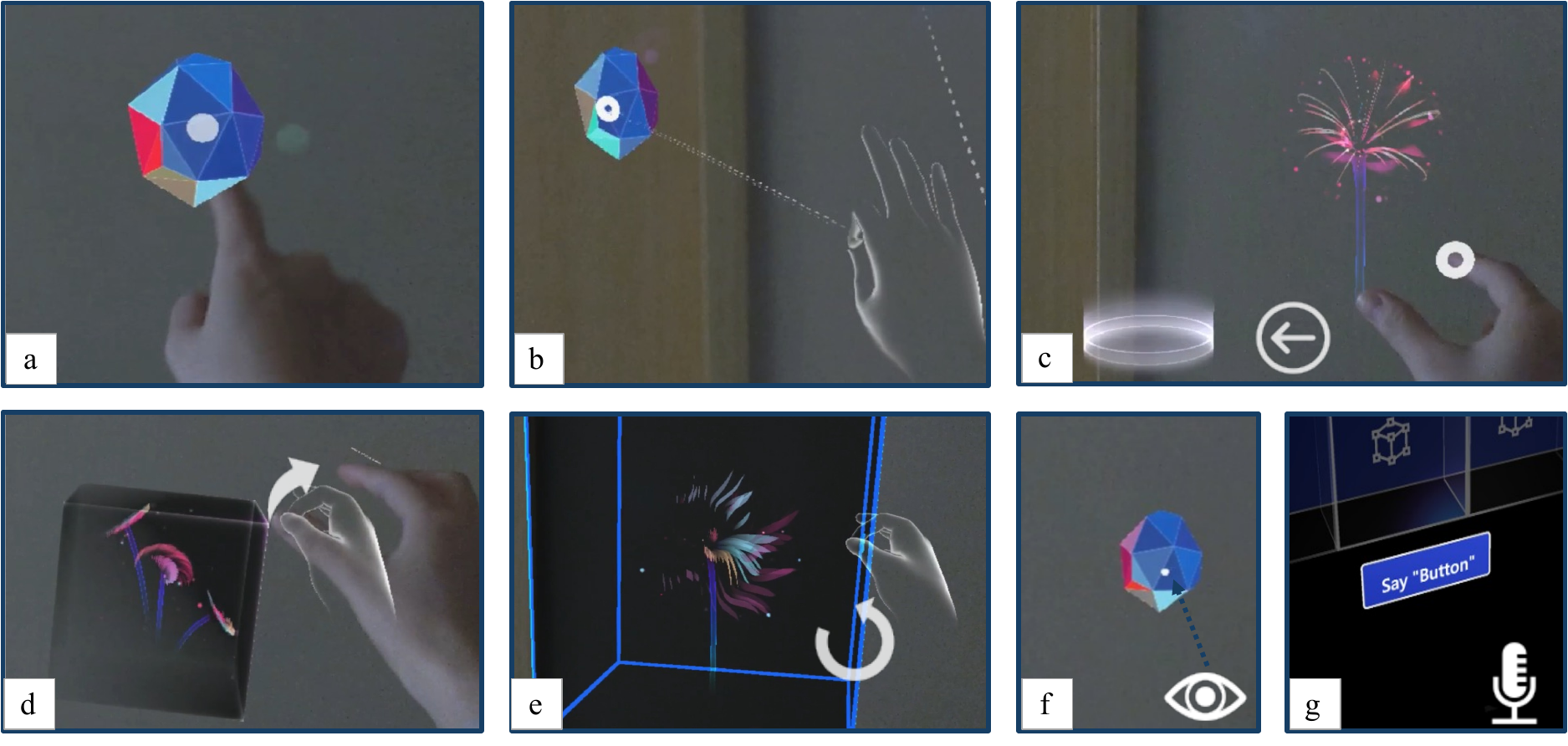}
\vspace{-0.2cm}
\caption{Diverse Interaction Modalities in AR. (a) Touch, (b) Virtual ray pointer, (c) Pinch and move, (d) Pinch and drag, (e) Pinch and rotate, (f) Eye gaze interaction, (g) Voice command.}
\Description{(a) depicts direct touch interaction using a hand. (b) showcases the use of a virtual ray pointer to pinpoint and select an object. (c) illustrates pinching a flower and subsequently moving it to a designated circle. (d) presents pinching a cube's vertex and dragging it for enlargement. (e) conveys pinching an edge of a cube to enact rotation. (f) highlights the utility of eye gaze for object selection, and (g) demonstrates vocal command as an interaction method.}
\label{fig:interaction}
\end{figure*}

\begin{figure*}[]
\includegraphics[width=5.7 in]{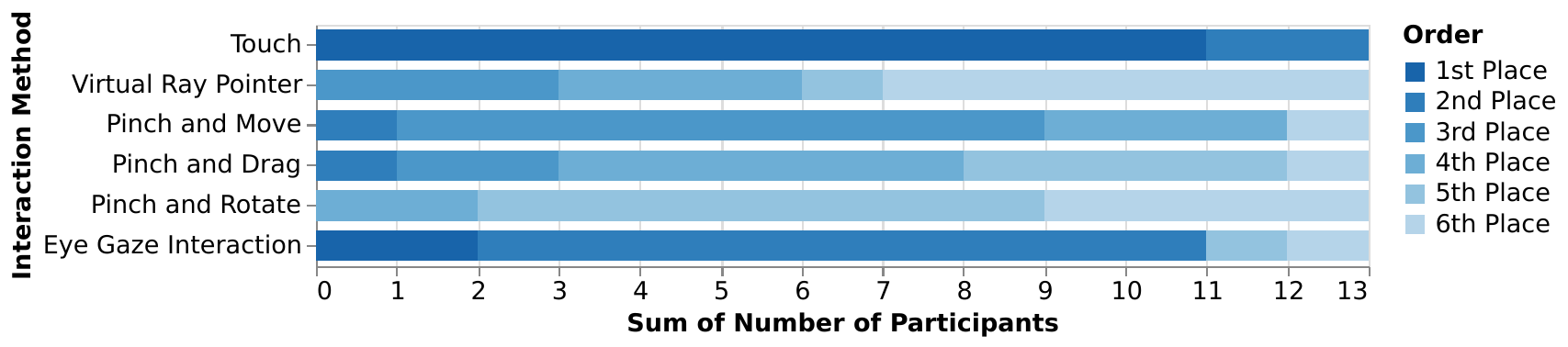}
\vspace{-0.2cm}
\caption{Participant Preferences for Interaction Methods. The y-axis delineates various interaction techniques while the x-axis quantifies the number of participants. For instance, the first place segment of the legend indicates the count of participants who ranked a specific interaction as their top preference.}
\Description{The chart is a colored bar chart. Different shades represent distinct ranking orders, with deeper hues signifying a higher preference by participants.}
\label{fig:order}
\end{figure*}

In examining how older adults interact with the Microsoft HoloLens 2 AR technology, participants experienced the seven core interaction methods of AR as shown in Figure~\ref{fig:interaction}. We aimed to identify what features of AR interaction were valued by older adults. We first let them express the pros and cons of each interaction method and asked them to rank these methods according to their preferences. For speech interaction, most of them struggled because they considered it more like a conversation function instead of issuing commands with a ``single operation'' like selecting. Therefore, we asked them to rank the remaining 6 interactions (Figure ~\ref{fig:order}). Except for two participants who did not complete all the interactions successfully, 13 participants ranked the six core interaction methods. In this section, we first report our participants' general perception of basic AR interactions and their detailed feedback for each interaction method. 

Direct touch interaction was most favored due to its similarity to physical interactions, followed by eye gaze interaction for its intuitive and effortless nature. Conversely, challenges arose with the virtual ray pointer and pinch-and-rotate functions, often due to recognition issues and operational difficulties. While the distinction between ``moving'' and ``dragging'' was seen as minimal, precision in pinching the vertex was a common challenge. We further summarize their feedback on each interaction method as follows. 

\noindent\textbf{Touch.} The instinctive nature of touch-based interactions, honed through years of smartphone use, made it the favorite AR interaction method among older participants. However, the intangibility of mid-air interactions posed a challenge. P9 expressed a longing for the solid, responsive feel of tangible objects, \textit{``There's a need to reach through to touch, and the depth perception isn't great. It feels unnatural, touching actual objects feels better.''} While P12 used a book near her to illustrate the tangible surfaces, explaining that it provided a more solid feeling, otherwise, she only relied on auditory feedback to confirm successful interaction. This longing for tangibility also resonated in P14's wish for content to manifest on walls, \textit{``the whole wall is like a big iPad''}. This suggests a desire among participants for tangible touch experiences that integrate seamlessly with their environment, such as automatically recognizing flat surfaces and giving users autonomy in determining the augmentation placement. This manifested in suggestions like using real-world tactile gestures, as showcased by P15's idea of table tapping for confirmation. She noted the tangible ``knocking'' or ``confirmation'' sensation it provided and appreciated not having to keep her arm elevated continuously.

\noindent\textbf{Virtual Ray Pointer.} 
Participants recognized it as an approach to interacting with far objects but found it hard to target closeby objects using the virtual ray pointer. P5 suggested using a handle would be helpful as he felt tired and experienced jitter when using his hand to point. However, our participants felt that it was unreasonable to use a controller to operate AR contents and put aside the controller to operate on the smartphone.

\noindent\textbf{Pinch and Move.} 
Two of our participants, P7 and P9, used wheelchairs. We observed that they faced difficulties adjusting their distance from the augmented content. Furthermore, ``pinch and move'' interactions were strenuous for them, particularly when maintaining consistent gestures over more extensive movements. Moreover, P15 expressed reservations about excessive physical movement due to safety concerns.
Acknowledging the decline in certain physical abilities among older adults, we collaborated with participants to conceptualize an interaction solution tailored to managing objects at varying distances. For distant features, an alternative method emerged: leveraging eye-tracking for distant interactions and employing simplistic hand gestures for closer AR content.

\noindent\textbf{Pinch and Drag.} 
The trials with the ``pinch'' gesture in AR revealed significant challenges, particularly around the precision demanded by the system. As P10 states, \textit{``The point, which is the finger's position. It is too narrow [for recognition] and felt unresponsive. We can not make it at that precise. If the goal is precision, it should differentiate various directions rather than restricting within such narrow bounds.''} Their proposed workaround---tracing an air circle for broader selection---emphasizes a desire for interactions with higher error tolerance. This approach, despite requiring more steps, imparts a sense of accomplishment and boosts confidence. For drag operations, some participants asked why not use the smartphone norm of two-finger resizing. Although designers might envision certain AR gestures as ``natural,'' interestingly, participants' comfort was rooted in familiarity, especially when standard smartphone gestures required less movement and effort.

\noindent\textbf{Pinch and Rotate.} Participants felt that this operation was challenging because it needed their wrists to flexibly rotate to a certain degree. Several participants found it hard to maintain the same gestures to rotate. 

\noindent\textbf{Eye Gaze Interaction.} Participants reported that they felt this interaction was natural. Although they had not used it before, they did not experience a learning barrier. However, for the confirmation of the selection, the default ``gaze'' did not work. Participants reported that in smartphone usage, sometimes they just needed more time to read or understand words, but this led to the accidental triggering of the gaze selection. They suggested using a phrase like ``yes, yes, yes'' for confirmation, just as they talked with themselves subconsciously using smartphones.

\noindent\textbf{Speech.} For the standard voice interaction usage shown in the demo for selecting, particularly in the context of learning smartphone applications, participants found it burdensome. P2 highlighted the exhausting nature of verbalizing every operation on a smartphone, saying, \textit{``Describing every single operation vocally would be exhausting. Using a smartphone requires numerous actions, and saying each one out loud would make me thirsty.''} Instead of utilizing voice for every selection, participants expressed a preference for more complex inquiries or dialogue-driven interactions. The consensus was that voice should serve as a supplementary or alternative mode, rather than the primary means of interaction. Participants also raised valid concerns about the feasibility of voice interactions in public settings. Background noise could hinder its effectiveness, and there was a shared apprehension about appearing weird when speaking aloud to devices in public spaces. Notably, participants highlighted that learning could occur in various environments, including outdoor settings.

\subsection{Older Adults' Concerns with AR for Supporting Smartphone Apps Exploring}
Although participants felt AR was interesting and helpful, they also raised concerns about price, weight and comfort, and ease of donning/doffing. The most frequently asked question was about the AR device's price. Participants felt that the price was too high for personal purchase (e.g., \textit{``It is useful but too expensive, I may not buy one at home if someone buys one in our older adults' center, I would like to use it there.''} -P7) Another concern was the heavy weight and discomfort after prolonged use. Several participants, particularly those with pre-existing neck conditions, felt the device's weight led to discomfort or strain on their neck, especially when thinking of exploring smartphone apps for a long time. This underscores the importance of ergonomics and lightweight designs in AR device development, especially when considering older adults as the target user group. P10 stated, \textit{``I am also concerned about other physical discomforts, like eye strain or dizziness, due to prolonged AR use.''} Additionally, participants highlighted the inconvenience of repeatedly donning and doffing AR devices when they wished to use them for exploring smartphone apps. P12 articulated a potential solution, suggesting, \textit{``If AR could be integrated into our glasses with a simple on-off switch, it would greatly enhance the convenience.''} P11 added that, especially in outdoor scenarios, carrying the device is not convenient and safe. 
                      
\section{Discussion}

We investigated how AR could be leveraged and designed to facilitate their smartphone app exploring experience. Through the tech-probe co-design workshop, our findings show that 1) AR can help reduce older adults' physical burden of exploring smartphone apps (e.g., reduced fine motor skills and tactile sensitivity; decreased visual acuity); 2) AR can help reduce older adults' cognitive burden of exploring smartphone apps (e.g., decreased working memory capacity, the decline in attentional resources); 3) AR can facilitate trial-and-error and mitigate mistake anxiety; 4) AR can help bridge the knowledge gap for older adults when exploring smartphone apps; 5) Older adults have their own preferences for interacting with AR (e.g., using a combination of tangible surfaces around them and space management); and 6) Older adults also raised three main concerns about adopting AR for app exploration: the expensive price for an at-home setting, physical discomforts such as neck strain, eye strain, dizziness, and burden of frequently donning/doffing AR devices.

\subsection{The Potentials of AR for Supporting Older Adults in Independent Learning}
AR offers an innovative approach to alleviate the challenges older adults face with smartphone interaction due to declining fine motor and visual skills. These individuals frequently miss small touch targets, leading to more errors, which has been substantiated by various studies~\cite{seidler2010,findlater2013age,jinsynapse2022,linwhy2018}. Though research has suggested usability improvements like adjusted icon sizes~\cite{targetleito2012, menghihow2017}, our participants found that enlarging content often compromised screen layout and reduced available preview information, a sentiment mirrored in earlier studies~\cite{ZIEFLE2010719}. In response, researchers have delved into automatic adjustments to enhance user experience~\cite{jinsynapse2022, shao2023smartphone}. AR, with its capability to project expansive displays, offers a compelling remedy. Our participants mentioned proportionally magnifying the screen in mid-air, removing the need for large physical screens, and ensuring adequate spacing between icons, which also minimizes touch errors. Additionally, they suggested using AR to simplify intricate interactions: instead of swiping, users see all apps and can intuitively tap to switch, which was the preferred gesture among older adults~\cite{menghihow2017, kobayashielderly2011}. Notably, AR addresses the visual challenges faced by older adults~\cite{czaja2006factors, o'brien2012}. Beyond the conventional method of simply enlarging content, our participants expressed a desire for a variety of magnifying techniques. These include dynamic presentations such as a pronounced brush style, a ``jump out'' effect, and multi-level zooming capabilities. Such features cater to the challenges posed by age-related visual acuity decline, ensuring a more tailored and enhanced user experience. \cm{We anticipate that by adapting mapping mechanisms from smartphones to AR, older adults with limitations in motor and visual skills can gain greater freedom and encouragement to explore more extensively.}

AR offers a distinctive means of addressing the challenges confronted by older adults, particularly when executing complex smartphone tasks against the backdrop of declining working memory and attention spans. As older adults grapple with tasks like transferring data between apps, their ability to remember specific app functionalities tends to wane~\cite{salthouse1994, Jin22I, guoolder2017}. Our study's participants highlighted AR's vast available visual space, which can encapsulate user actions, illustrate processes, and synthesize data from disparate apps to aid memory. Moreover, with the myriad of notifications, pop-ups, and multimedia diversions on smartphones, older users find their already limited attention further dispersed~\cite{craik1986}. Suggestions from our participants include AR's potential to emphasize ongoing tasks via an extensive reminder board. \cm{Leveraging AR as an additional memory-aid learning tool is feasible, with the potential to reduce cognitive burden and encourage easier exploration.}          

As for supporting older adults in learning to use smartphone apps, past research has ventured into streamlining these interactions using strategies like step-by-step guides~\cite{leunghow2012, Pang2021Technology}, illustrative video demonstrations~\cite{cyartoactive2016}, and intricate augmented display systems~\cite{Wilson2018Help}. A limitation with many of these interventions lies in their static designs, potentially hindering the exploratory, trial-and-error learning method many older adults favor~\cite{Pang2021Technology}. An exemplar in this space, Jin et al.'s system, advocates for this learning mode by retaining an operational state to reduce ambiguities~\cite{jinsynapse2022}. Nevertheless, doubts linger over its effectiveness, especially when users persistently choose incorrect options~\cite{mahmud2020learning}. Bridging these endeavors and the untapped potential of newer technologies, AR use vast visual environment sets it apart, vividly capturing user interactions to bolster both real-time participation and the trial-and-error approach. \cm{Enabling cross-device communication allows AR to capture smartphone events and efficiently generate exploratory maps, thus enhancing the exploratory journey.} This paradigm does not just offer immediate visual responses but also facilitates storing several operational states as a backtracking mechanism. When users approach final decisions, a visual timeline of their entire journey provides an added confidence layer, particularly when navigating unfamiliar smartphone cues or fearing inadvertent actions{~\cite{sinha2013, mitzner2010older}.}

AR's rich visual palette can enhance the elucidation of complex terminologies encountered during exploration, showing potential to bolster learning outcomes~\cite{albrecht2013effects,huang2021adaptutar}. Previous research has highlighted the efficacy of multimedia-based instructions for older adults~\cite{VanGerven2003}, and AR, as an emerging medium, brings additional value, particularly through its spatial capabilities. \cm{The integration of advanced large language models (LLMs) and AI-generated content (AIGC) tools significantly enhances the ability to provide relevant contextual knowledge and explanations.} Our participants not only expressed a desire for diverse modes of presentation but were also captivated by AR's immersive spatial characteristics. Such engagement can kindle their curiosity, prompting them to delve deeper into smartphone applications. Furthermore, with AR being a separate device designed with security in mind, participants expressed fewer apprehensions compared to other applications on their phones, mitigating potential security concerns~\cite{Jin22I,jinsynapse2022, Akter2023It}.

\subsection{Improving AR Interactions to Better Support Mobile App Exploration Among Older Adults}
\label{section:interaction}

One of the current challenges in AR development is the lack of guidelines including interaction~\cite{ashtari2020creating}. Our study stands on older adults' perceptive and provides valuable insights for future interaction design to be more inclusive.

\textbf{AR interactions should be similar to the reality of using smartphones.} Participants exhibited a clear inclination for AR interactions that mirrored their routine smartphone habits since tapping is the easiest gesture for older people~\cite{menghihow2017, kobayashielderly2011}. Participants favored AR interactions, such as ``touch'' and ``eye gaze'', wanted to use their finger as a magnifying tool, and preferred the familiar two-finger smartphone gesture to enlarge, which mirrored established smartphone practices. This echoes Norman's comment that natural user interfaces are not natural~\cite{norman2010natural}: participants gravitated more toward interactions reminiscent of smartphone usage rather than ones perceived as innately natural.



\textbf{Voice interactions should act as a complementary input method, not a primary mechanism.} Voice interaction, while promising, showed limitations in the context of learning smartphone apps. Participants found it cumbersome to use voice commands for every individual operation, which may not be useful in outdoor scenarios due to ambient noise and social awkwardness. Participants endorsed voice interactions as supplementary, succinctly affirming gestures. This sentiment resonates with prior studies emphasizing the benefits of amalgamating voice with gestures ~\cite{irawati2006evaluation,lee2008wizard, piumsomboon2014grasp}.



\textbf{Interactions should have more error tolerance and provide psychological comfort.} Precision in AR proved to be a challenge for many older participants. They preferred having a broader, more encompassing interaction followed by refinement. Interestingly, this was not purely about operational efficiency; it catered to their psychological comfort, offering a sense of achievement.



\textbf{AR interactions should involve tangible surfaces that are anchored in the physical world.} Participants’ feedback underscored the need to bridge the tangible and digital in AR interfaces. The act of virtually extending one's hand into the void, without tangible feedback was perceived as counterintuitive ~\cite{bruder2013touching}. This explains older adults' inclination towards a blend of digital immersion and tangible references. For these users, AR systems that can intuitively identify and adapt to proximate surfaces could significantly improve the experience by offering familiar interaction touchpoints. Participants, for instance, cited books, tables, and walls as tangible anchors. This tangible engagement could also offer ergonomic benefits, potentially alleviating the need for sustained postures that could strain users. The evolving realm of tangible AR~\cite{billinghurst2008tangible}, which leverages physical objects as conduits or anchors for digital interplay, emerges as a potential solution. Researchers have developed techniques that enable virtually any surface to double as a touch interface are indicative of this trend~\cite{xiao2018mrtouch} and tailored edges on commonplace items for tactile feedback and easy tracking~\cite{he2023ubi}. It is imperative for future tangible AR research to acknowledge and cater to older adults, framing them as a core user demographic.


\textbf{AR interactions for older adults should consider spatial orientations that promote better posture and health.} The inherent spatial adaptability of AR offers a diverse array of content positions, each with distinct health and engagement implications. While prior research has explored augmenting additional digital content around smartphones~\cite{reichherzer2021secondsight,bang2020effects,zhu2020bishare,hubenschmid2023around}, this study brings the preferences of elderly users into focus. Our findings indicate a strong preference among older adults for horizontal AR content placements, favoring the left or right for its immediate accessibility. Interestingly, vertical top-placed layouts carried potential health dividends. Encouraging users to periodically elevate their gaze might counteract the downward smartphone posture, offering relief against potential neck strain. While these insights shed light on older users' needs, it's vital to contextualize them within broader AR research. Many extant studies, like that by Hubenschmid et al.~\cite{hubenschmid2023around} which probed AR space dimensions around smartphones, predominantly cater to younger users. Their assertion---magnifying AR space akin to a desktop screen optimizes spatial memory and user experience---may not resonate universally, especially given the potentially limited tech familiarity among older adults. As the AR realm evolves, integrating the needs and preferences of the elderly becomes a necessity.

\subsection{Improving the Adoption of AR Supporting Tools Among Older Adults}

\revision{Although we showed the potential of AR for supporting mobile app exploration, our findings also revealed seven factors that could affect the adoption of AR among older adults, which will be discussed next.} \revision{We will use the UTAUT (Unified Theory of Acceptance and Use of Technology) and UTAUT2 models~\cite{Venkatesh2003, Venkatesh2012} to help explain the factors.} 

\textbf{Device Physicality and Usability.} The portability and weight of AR devices emerged as significant concerns. While participants were keen on leveraging AR across diverse contexts, including indoors and outdoors, the current form factor impedes ubiquitous use. This observation is rooted in the effort expectancy facet of UTAUT~\cite{Venkatesh2003}. In response, future designs could focus on lighter AR headsets or even integrate AR functionalities into standard eyeglasses, promoting ease of use. \revision{In our work, we choose Hololens2, which has better presenting performance and similar wight to the device for older adults in previous work~\cite{Mostajeran2020Augmented}. A more recent AR device apple vision pro~\cite{AppleVisionPro} declares that lighter and better using experience and we believe that researchers have been putting efforts in improving the devices. Our findings further motivate them to care about the weights and convenience of taking on and off.}

\textbf{Economic Considerations.}
The price of the device seemed unjustifiable if its primary use case was aiding smartphone interactions. This resonates with the UTAUT2's price value determinant~\cite{Venkatesh2012}. Hence, introducing AR tools in communal settings like older adult centers might be an interim solution. This would also address facilitating conditions, proposed in UTAUT~\cite{Venkatesh2003}, ensuring regular maintenance by dedicated personnel. \revision{Moreover, producers should further consider using affordable materials and reducing the cost of the device to make it more acceptable for more users.}

\textbf{\revision{Social} Learning in AR.}
Our study echoes prior research advocating for collaborative AR experiences~\cite{Bai2020Auser, Villanueva2020Meta, Schlagowski2023Wish}. With older adults expressing a preference for joint exploration, it opens avenues for co-learning in AR environments \revision{either for the social interaction between the helper and the older adults or among older adults themselves. Beyond physically sharing the social learning environment, remote social learning~\cite{ Schlagowski2023Wish} also deserves exploration to make older adults learn more flexible.} Such initiatives align with the UTAUT's social influence dimension~\cite{Venkatesh2003}, underpinning the significance of shared technological experiences.

\textbf{Security and Privacy.}
AR's potential security threats, such as data leaks through camera access~\cite{Hadar2018Toward, Chen2018Acase, Gugenheimer2022Novel}, warrant attention. Participants' perception of AR as a distinct device from their smartphones did offer a semblance of safety. Yet, given the cross-device interactions, a stringent regulatory framework is indispensable. Techniques like disabling certain functions during smartphone access in AR can provide added layers of security.

\textbf{Health and Well-being.}
The health implications of prolonged AR usage, especially concerns like motion sickness among older adults, necessitate periodic usage checks~\cite{Lee2019Potential}. Intriguingly, integrating AR with healthcare applications, as alluded to by participants, could enhance smartphone usage and promote health concurrently. This aligns with UTAUT's performance expectancy, emphasizing tangible benefits from technology adoption~\cite{Venkatesh2003}.

\textbf{Gamification and Engagement.}
Our findings underscore the value of hedonic motivation of UTAUT2~\cite{Venkatesh2012}, with participants appreciating engaging AR features. Integrating AR with gamified experiences, perhaps even serious games to bolster cognitive functions~\cite{Han2011Mobile}, could serve dual purposes: aiding smartphone exploration and fostering cognitive enhancement. 

\textbf{Visibility of Accessibility Features.}
Our study highlights the need for accessibility features in AR since cognitive shifts in older adults could limit their inclination to explore new technology. Therefore, AR designs should prioritize accessibility features, ensuring they are intuitive and salient for the elderly demographic~\cite{Franz2019Perception}. A tangible design example might be the incorporation of an easily identifiable button that toggles these functions, simplifying their discovery and use.

\subsection{Limitation and Future Work}

Our study primarily centered on older adults within a specific geographical purview, potentially introducing cultural and regional biases. This nuances the generalizability of our findings, particularly as our cohort was technology-receptive. Consequently, the insights may not be representative of a wider demographic less inclined toward technological adoption. \revision{Future work should expand the sample size to include a more diverse population of older adults such as those who are less familiar with technology or who have different levels of cognitive or physical abilities. }

While our investigation delved into the potentialities of AR in aiding older adults to navigate smartphone applications, it positioned itself predominantly within the design realm. We did not conduct comparative evaluations to determine the exact efficacy of AR vs. a baseline in task accomplishments. In evaluating the nuances of AR assistance, future research can delineate the outcomes of AR-facilitated smartphone engagements \revision{based on this work's findings}. \revision{For example, including a control group in future studies would enable a more precise assessment of AR interventions' effectiveness on specific tasks, thereby illustrating the extent to which AR facilitates older adults' learning of digital technologies. Additionally, a longitudinal approach, involving extended follow-up periods, would be beneficial to understand the enduring effects of AR interventions on the app exploration capabilities and confidence of older adults.}

Additionally, our research scope was confined to solitary application explorations. Future endeavors might benefit from including collaborative AR scenarios. Such multi-user frameworks, as indicated by some participants, hold the promise of fostering shared learning landscapes. However, as we traverse this collaborative domain, researchers should take notice of ensuring the privacy and security of older users, which is critical in collaborative AR~\cite{Rajaram2023Eliciting}.


\section{Conclusion}
Amid the rise of the global aging trend, the imperative for older adults to adapt to a rapidly changing digital landscape is unsurpassed. This study elucidates AR's potential in simplifying this transition, particularly in the realm of smartphone application exploration. Through our two-phase study, we observed that AR can be pivotal in mitigating both physical and cognitive challenges that older adults encounter, offering a more intuitive and streamlined learning experience. Notably, during multi-app interactions and stages characterized by trial and error, AR's impact was especially pronounced. Our examination of older adults' interactional patterns with AR has also shed light on key design strategies, emphasizing the need for customization and a deeper understanding of their interaction preferences, thus further contributing to the future inclusive AR interaction design. We further underscore the foundational role of AR in crafting future tools specifically tailored for older adults, including the importance of weaving healthcare considerations within AR applications, delve into the potential of AR-facilitated collaborative learning for smartphone apps—integrating vital social dimensions—and champion the infusion of gamification techniques to amplify their engagement and interest.

\section{Acknowledgments}
We extend our gratitude to the Shanghai Pudong Biyun Art Gallery, Jinqiao Nursing House, and community staff for their collaboration, and to all participants for their valuable time. Special thanks to my friends at Vislab, Yanna and Rui, for their assistance with the figures. 
\bibliographystyle{ACM-Reference-Format}
\bibliography{main}


\begin{thebibliography}{104}


\ifx \showCODEN    \undefined \def \showCODEN     #1{\unskip}     \fi
\ifx \showDOI      \undefined \def \showDOI       #1{#1}\fi
\ifx \showISBNx    \undefined \def \showISBNx     #1{\unskip}     \fi
\ifx \showISBNxiii \undefined \def \showISBNxiii  #1{\unskip}     \fi
\ifx \showISSN     \undefined \def \showISSN      #1{\unskip}     \fi
\ifx \showLCCN     \undefined \def \showLCCN      #1{\unskip}     \fi
\ifx \shownote     \undefined \def \shownote      #1{#1}          \fi
\ifx \showarticletitle \undefined \def \showarticletitle #1{#1}   \fi
\ifx \showURL      \undefined \def \showURL       {\relax}        \fi
\providecommand\bibfield[2]{#2}
\providecommand\bibinfo[2]{#2}
\providecommand\natexlab[1]{#1}
\providecommand\showeprint[2][]{arXiv:#2}

\bibitem[cou(2021)]%
        {coursera}
 \bibinfo{year}{2021}\natexlab{}.
\newblock \bibinfo{title}{Coursera}.
\newblock \bibinfo{howpublished}{\url{https://www.coursera.org/}}.
\newblock
\newblock
\shownote{Accessed: 2023-09-07}.


\bibitem[edx(2021)]%
        {edx}
 \bibinfo{year}{2021}\natexlab{}.
\newblock \bibinfo{title}{edX}.
\newblock \bibinfo{howpublished}{\url{https://www.edx.org/}}.
\newblock
\newblock
\shownote{Accessed: 2023-09-07}.


\bibitem[uda(2021)]%
        {udacity}
 \bibinfo{year}{2021}\natexlab{}.
\newblock \bibinfo{title}{Udacity}.
\newblock \bibinfo{howpublished}{\url{https://www.udacity.com/}}.
\newblock
\newblock
\shownote{Accessed: 2023-09-07}.


\bibitem[Akter et~al\mbox{.}(2023)]%
        {Akter2023It}
\bibfield{author}{\bibinfo{person}{Mamtaj Akter}, \bibinfo{person}{Leena
  Alghamdi}, \bibinfo{person}{Jess Kropczynski},
  \bibinfo{person}{Heather~Richter Lipford}, {and} \bibinfo{person}{Pamela~J.
  Wisniewski}.} \bibinfo{year}{2023}\natexlab{}.
\newblock \showarticletitle{It Takes a Village: A Case for Including Extended
  Family Members in the Joint Oversight of Family-Based Privacy and Security
  for Mobile Smartphones}. In \bibinfo{booktitle}{\emph{Extended Abstracts of
  the 2023 CHI Conference on Human Factors in Computing Systems}} (Hamburg,
  Germany) \emph{(\bibinfo{series}{CHI EA '23})}.
  \bibinfo{publisher}{Association for Computing Machinery},
  \bibinfo{address}{New York, NY, USA}, Article \bibinfo{articleno}{194},
  \bibinfo{numpages}{7}~pages.
\newblock
\showISBNx{9781450394222}
\urldef\tempurl%
\url{https://doi.org/10.1145/3544549.3585904}
\showDOI{\tempurl}


\bibitem[Albrecht et~al\mbox{.}(2013)]%
        {albrecht2013effects}
\bibfield{author}{\bibinfo{person}{Urs-Vito Albrecht},
  \bibinfo{person}{Kristian Folta-Schoofs}, \bibinfo{person}{Marianne
  Behrends}, \bibinfo{person}{Ute Von~Jan}, {et~al\mbox{.}}}
  \bibinfo{year}{2013}\natexlab{}.
\newblock \showarticletitle{Effects of mobile augmented reality learning
  compared to textbook learning on medical students: randomized controlled
  pilot study}.
\newblock \bibinfo{journal}{\emph{Journal of medical Internet research}}
  \bibinfo{volume}{15}, \bibinfo{number}{8} (\bibinfo{year}{2013}),
  \bibinfo{pages}{e2497}.
\newblock


\bibitem[Apple(2023)]%
        {AppleVisionPro}
\bibfield{author}{\bibinfo{person}{Apple}.} \bibinfo{year}{2023}\natexlab{}.
\newblock \bibinfo{booktitle}{\emph{Apple Vision Pro - Apple}}.
\newblock
\urldef\tempurl%
\url{https://www.apple.com/apple-vision-pro/}
\showURL{%
\tempurl}
\newblock
\shownote{Accessed: 2023-11-28}.


\bibitem[Ashtari et~al\mbox{.}(2020)]%
        {ashtari2020creating}
\bibfield{author}{\bibinfo{person}{Narges Ashtari}, \bibinfo{person}{Andrea
  Bunt}, \bibinfo{person}{Joanna McGrenere}, \bibinfo{person}{Michael
  Nebeling}, {and} \bibinfo{person}{Parmit~K Chilana}.}
  \bibinfo{year}{2020}\natexlab{}.
\newblock \showarticletitle{Creating augmented and virtual reality
  applications: Current practices, challenges, and opportunities}. In
  \bibinfo{booktitle}{\emph{Proceedings of the 2020 CHI conference on human
  factors in computing systems}}. \bibinfo{pages}{1--13}.
\newblock


\bibitem[Atkinson et~al\mbox{.}(2016)]%
        {atkinsonbreaking2016}
\bibfield{author}{\bibinfo{person}{Keith Atkinson}, \bibinfo{person}{Jaclyn
  Barnes}, \bibinfo{person}{Judith Albee}, \bibinfo{person}{Peter Anttila},
  \bibinfo{person}{Judith Haataja}, \bibinfo{person}{Kanak Nanavati},
  \bibinfo{person}{Kelly Steelman}, {and} \bibinfo{person}{Charles Wallace}.}
  \bibinfo{year}{2016}\natexlab{}.
\newblock \showarticletitle{Breaking {Barriers} to {Digital} {Literacy}: {An}
  {Intergenerational} {Social}-{Cognitive} {Approach}}. In
  \bibinfo{booktitle}{\emph{Proceedings of the 18th {International} {ACM}
  {SIGACCESS} {Conference} on {Computers} and {Accessibility}}}.
  \bibinfo{publisher}{ACM}, \bibinfo{address}{Reno Nevada USA},
  \bibinfo{pages}{239--244}.
\newblock
\showISBNx{978-1-4503-4124-0}
\urldef\tempurl%
\url{https://doi.org/10.1145/2982142.2982183}
\showDOI{\tempurl}


\bibitem[Bai et~al\mbox{.}(2020)]%
        {Bai2020Auser}
\bibfield{author}{\bibinfo{person}{Huidong Bai}, \bibinfo{person}{Prasanth
  Sasikumar}, \bibinfo{person}{Jing Yang}, {and} \bibinfo{person}{Mark
  Billinghurst}.} \bibinfo{year}{2020}\natexlab{}.
\newblock \showarticletitle{A User Study on Mixed Reality Remote Collaboration
  with Eye Gaze and Hand Gesture Sharing}. In
  \bibinfo{booktitle}{\emph{Proceedings of the 2020 CHI Conference on Human
  Factors in Computing Systems}} (Honolulu, HI, USA)
  \emph{(\bibinfo{series}{CHI '20})}. \bibinfo{publisher}{Association for
  Computing Machinery}, \bibinfo{address}{New York, NY, USA},
  \bibinfo{pages}{1–13}.
\newblock
\showISBNx{9781450367080}
\urldef\tempurl%
\url{https://doi.org/10.1145/3313831.3376550}
\showDOI{\tempurl}


\bibitem[Bang et~al\mbox{.}(2020)]%
        {bang2020effects}
\bibfield{author}{\bibinfo{person}{Sunyoung Bang}, \bibinfo{person}{Hyunjin
  Lee}, {and} \bibinfo{person}{Woontack Woo}.} \bibinfo{year}{2020}\natexlab{}.
\newblock \showarticletitle{Effects of augmented content’s placement and size
  on user’s search experience in extended displays}. In
  \bibinfo{booktitle}{\emph{2020 IEEE International Symposium on Mixed and
  Augmented Reality Adjunct (ISMAR-Adjunct)}}. IEEE, \bibinfo{pages}{184--188}.
\newblock


\bibitem[Barnard et~al\mbox{.}(2013)]%
        {barnard2013learning}
\bibfield{author}{\bibinfo{person}{Yvonne Barnard}, \bibinfo{person}{Mike~D
  Bradley}, \bibinfo{person}{Frances Hodgson}, {and} \bibinfo{person}{Ashley~D
  Lloyd}.} \bibinfo{year}{2013}\natexlab{}.
\newblock \showarticletitle{Learning to use new technologies by older adults:
  Perceived difficulties, experimentation behaviour and usability}.
\newblock \bibinfo{journal}{\emph{Computers in human behavior}}
  \bibinfo{volume}{29}, \bibinfo{number}{4} (\bibinfo{year}{2013}),
  \bibinfo{pages}{1715--1724}.
\newblock


\bibitem[Beh et~al\mbox{.}(2015)]%
        {behwhere2015}
\bibfield{author}{\bibinfo{person}{Jeanie Beh}, \bibinfo{person}{Sonja Pedell},
  {and} \bibinfo{person}{Wendy Doube}.} \bibinfo{year}{2015}\natexlab{}.
\newblock \showarticletitle{Where is the "{I}" in {iPad}?: {The} {Role} of
  {Interest} in {Older} {Adults}' {Learning} of {Mobile} {Touch} {Screen}
  {Technologies}}. In \bibinfo{booktitle}{\emph{Proceedings of the {Annual}
  {Meeting} of the {Australian} {Special} {Interest} {Group} for {Computer}
  {Human} {Interaction}}}. \bibinfo{publisher}{ACM},
  \bibinfo{address}{Parkville VIC Australia}, \bibinfo{pages}{437--445}.
\newblock
\showISBNx{978-1-4503-3673-4}
\urldef\tempurl%
\url{https://doi.org/10.1145/2838739.2838776}
\showDOI{\tempurl}


\bibitem[Berenguer et~al\mbox{.}(2017)]%
        {BerenguerAnabela2017ASUA}
\bibfield{author}{\bibinfo{person}{Anabela Berenguer}, \bibinfo{person}{Jorge
  Goncalves}, \bibinfo{person}{Simo Hosio}, \bibinfo{person}{Denzil Ferreira},
  \bibinfo{person}{Theodoros Anagnostopoulos}, {and} \bibinfo{person}{Vassilis
  Kostakos}.} \bibinfo{year}{2017}\natexlab{}.
\newblock \showarticletitle{Are Smartphones Ubiquitous?: An in-depth survey of
  smartphone adoption by seniors}.
\newblock \bibinfo{journal}{\emph{IEEE consumer electronics magazine}}
  \bibinfo{volume}{6}, \bibinfo{number}{1} (\bibinfo{year}{2017}),
  \bibinfo{pages}{104--110}.
\newblock
\showISSN{2162-2248}


\bibitem[Bianco et~al\mbox{.}(2016)]%
        {Bianco2016Augmented}
\bibfield{author}{\bibinfo{person}{Michael~Lo Bianco}, \bibinfo{person}{Sonja
  Pedell}, {and} \bibinfo{person}{Gianni Renda}.}
  \bibinfo{year}{2016}\natexlab{}.
\newblock \showarticletitle{Augmented Reality and Home Modifications: A Tool to
  Empower Older Adults in Fall Prevention}. In
  \bibinfo{booktitle}{\emph{Proceedings of the 28th Australian Conference on
  Computer-Human Interaction}} (Launceston, Tasmania, Australia)
  \emph{(\bibinfo{series}{OzCHI '16})}. \bibinfo{publisher}{Association for
  Computing Machinery}, \bibinfo{address}{New York, NY, USA},
  \bibinfo{pages}{499–507}.
\newblock
\showISBNx{9781450346184}
\urldef\tempurl%
\url{https://doi.org/10.1145/3010915.3010929}
\showDOI{\tempurl}


\bibitem[Billinghurst et~al\mbox{.}(2008)]%
        {billinghurst2008tangible}
\bibfield{author}{\bibinfo{person}{Mark Billinghurst},
  \bibinfo{person}{Hirokazu Kato}, \bibinfo{person}{Ivan Poupyrev},
  {et~al\mbox{.}}} \bibinfo{year}{2008}\natexlab{}.
\newblock \showarticletitle{Tangible augmented reality}.
\newblock \bibinfo{journal}{\emph{Acm siggraph asia}} \bibinfo{volume}{7},
  \bibinfo{number}{2} (\bibinfo{year}{2008}), \bibinfo{pages}{1--10}.
\newblock


\bibitem[Bruder et~al\mbox{.}(2013)]%
        {bruder2013touching}
\bibfield{author}{\bibinfo{person}{Gerd Bruder}, \bibinfo{person}{Frank
  Steinicke}, {and} \bibinfo{person}{Wolfgang Stuerzlinger}.}
  \bibinfo{year}{2013}\natexlab{}.
\newblock \showarticletitle{Touching the void revisited: Analyses of touch
  behavior on and above tabletop surfaces}. In
  \bibinfo{booktitle}{\emph{Human-Computer Interaction--INTERACT 2013: 14th
  IFIP TC 13 International Conference, Cape Town, South Africa, September 2-6,
  2013, Proceedings, Part I 14}}. Springer, \bibinfo{pages}{278--296}.
\newblock


\bibitem[Center(2022)]%
        {pew2022techusers}
\bibfield{author}{\bibinfo{person}{Pew~Research Center}.}
  \bibinfo{year}{2022}\natexlab{}.
\newblock \showarticletitle{Share of those 65 and older who are tech users has
  grown in the past decade}.
\newblock \bibinfo{journal}{\emph{Pew Research Center}} (\bibinfo{date}{13 Jan}
  \bibinfo{year}{2022}).
\newblock
\urldef\tempurl%
\url{https://www.pewresearch.org/short-reads/2022/01/13/share-of-those-65-and-older-who-are-tech-users-has-grown-in-the-past-decade/}
\showURL{%
\tempurl}
\newblock
\shownote{Accessed: [2023-9-13]}.


\bibitem[Cerna et~al\mbox{.}(2022)]%
        {cernasituated2022}
\bibfield{author}{\bibinfo{person}{*~Katerina Cerna}, \bibinfo{person}{Claudia
  Müller}, \bibinfo{person}{Dave Randall}, {and} \bibinfo{person}{Martin
  Hunker}.} \bibinfo{year}{2022}\natexlab{}.
\newblock \showarticletitle{Situated {Scaffolding} for {Sustainable}
  {Participatory} {Design}: {Learning} {Online} with {Older} {Adults}}.
\newblock \bibinfo{journal}{\emph{Proceedings of the ACM on Human-Computer
  Interaction}} \bibinfo{volume}{6}, \bibinfo{number}{GROUP}
  (\bibinfo{date}{Jan.} \bibinfo{year}{2022}), \bibinfo{pages}{1--25}.
\newblock
\showISSN{2573-0142}
\urldef\tempurl%
\url{https://doi.org/10.1145/3492831}
\showDOI{\tempurl}


\bibitem[Chen et~al\mbox{.}(2018)]%
        {Chen2018Acase}
\bibfield{author}{\bibinfo{person}{Song Chen}, \bibinfo{person}{Zupei Li},
  \bibinfo{person}{Fabrizio Dangelo}, \bibinfo{person}{Chao Gao}, {and}
  \bibinfo{person}{Xinwen Fu}.} \bibinfo{year}{2018}\natexlab{}.
\newblock \showarticletitle{A Case Study of Security and Privacy Threats from
  Augmented Reality (AR)}. In \bibinfo{booktitle}{\emph{2018 International
  Conference on Computing, Networking and Communications (ICNC)}}.
  \bibinfo{pages}{442--446}.
\newblock
\urldef\tempurl%
\url{https://doi.org/10.1109/ICCNC.2018.8390291}
\showDOI{\tempurl}


\bibitem[Chin and Fu(2012)]%
        {chinage2012}
\bibfield{author}{\bibinfo{person}{Jessie Chin} {and} \bibinfo{person}{Wai-Tat
  Fu}.} \bibinfo{year}{2012}\natexlab{}.
\newblock \showarticletitle{Age differences in exploratory learning from a
  health information website}. In \bibinfo{booktitle}{\emph{Proceedings of the
  {SIGCHI} {Conference} on {Human} {Factors} in {Computing} {Systems}}}.
  \bibinfo{publisher}{ACM}, \bibinfo{address}{Austin Texas USA},
  \bibinfo{pages}{3031--3040}.
\newblock
\showISBNx{978-1-4503-1015-4}
\urldef\tempurl%
\url{https://doi.org/10.1145/2207676.2208715}
\showDOI{\tempurl}


\bibitem[Choudrie et~al\mbox{.}(2018)]%
        {Choudrive2018}
\bibfield{author}{\bibinfo{person}{Jyoti Choudrie},
  \bibinfo{person}{Chike~Obuekwe Junior}, \bibinfo{person}{Brad McKenna}, {and}
  \bibinfo{person}{Shahper Richter}.} \bibinfo{year}{2018}\natexlab{}.
\newblock \showarticletitle{Understanding and conceptualising the adoption, use
  and diffusion of mobile banking in older adults: A research agenda and
  conceptual framework}.
\newblock \bibinfo{journal}{\emph{Journal of Business Research}}
  \bibinfo{volume}{88} (\bibinfo{year}{2018}), \bibinfo{pages}{449--465}.
\newblock


\bibitem[Conte and Munteanu(2019)]%
        {contehelp2019}
\bibfield{author}{\bibinfo{person}{Sho Conte} {and} \bibinfo{person}{Cosmin
  Munteanu}.} \bibinfo{year}{2019}\natexlab{}.
\newblock \showarticletitle{Help!: {I}'m {Stuck}, and there's no {F1} {Key} on
  {My} {Tablet}!}. In \bibinfo{booktitle}{\emph{Proceedings of the 21st
  {International} {Conference} on {Human}-{Computer} {Interaction} with
  {Mobile} {Devices} and {Services}}}. \bibinfo{publisher}{ACM},
  \bibinfo{address}{Taipei Taiwan}, \bibinfo{pages}{1--11}.
\newblock
\showISBNx{978-1-4503-6825-4}
\urldef\tempurl%
\url{https://doi.org/10.1145/3338286.3340121}
\showDOI{\tempurl}


\bibitem[Corbin and Strauss(2014)]%
        {corbin2014basics}
\bibfield{author}{\bibinfo{person}{Juliet Corbin} {and} \bibinfo{person}{Anselm
  Strauss}.} \bibinfo{year}{2014}\natexlab{}.
\newblock \bibinfo{booktitle}{\emph{Basics of Qualitative Research: Techniques
  and Procedures for Developing Grounded Theory}}.
\newblock \bibinfo{publisher}{Sage Publications}.
\newblock


\bibitem[Craik(1986)]%
        {craik1986}
\bibfield{author}{\bibinfo{person}{F.I.M. Craik}.}
  \bibinfo{year}{1986}\natexlab{}.
\newblock \showarticletitle{A functional account of age differences in memory}.
\newblock  (\bibinfo{year}{1986}), \bibinfo{pages}{409--422}.
\newblock


\bibitem[Cruz-Neira et~al\mbox{.}(1992)]%
        {cruz1992cave}
\bibfield{author}{\bibinfo{person}{Carolina Cruz-Neira},
  \bibinfo{person}{Daniel~J Sandin}, \bibinfo{person}{Thomas~A DeFanti},
  \bibinfo{person}{Robert~V Kenyon}, {and} \bibinfo{person}{John~C Hart}.}
  \bibinfo{year}{1992}\natexlab{}.
\newblock \showarticletitle{The CAVE: audio visual experience automatic virtual
  environment}.
\newblock \bibinfo{journal}{\emph{Commun. ACM}} \bibinfo{volume}{35},
  \bibinfo{number}{6} (\bibinfo{year}{1992}), \bibinfo{pages}{64--73}.
\newblock


\bibitem[Cyarto et~al\mbox{.}(2016)]%
        {cyartoactive2016}
\bibfield{author}{\bibinfo{person}{Elizabeth~V. Cyarto},
  \bibinfo{person}{Frances Batchelor}, \bibinfo{person}{Steven Baker}, {and}
  \bibinfo{person}{Briony Dow}.} \bibinfo{year}{2016}\natexlab{}.
\newblock \showarticletitle{Active ageing with avatars: a virtual exercise
  class for older adults}. In \bibinfo{booktitle}{\emph{Proceedings of the 28th
  {Australian} {Conference} on {Computer}-{Human} {Interaction} - {OzCHI}
  '16}}. \bibinfo{publisher}{ACM Press}, \bibinfo{address}{Launceston,
  Tasmania, Australia}, \bibinfo{pages}{302--309}.
\newblock
\showISBNx{978-1-4503-4618-4}
\urldef\tempurl%
\url{https://doi.org/10.1145/3010915.3010944}
\showDOI{\tempurl}


\bibitem[Czaja et~al\mbox{.}(2006)]%
        {czaja2006factors}
\bibfield{author}{\bibinfo{person}{Sara~J Czaja}, \bibinfo{person}{Neil
  Charness}, \bibinfo{person}{Arthur~D Fisk}, \bibinfo{person}{Christopher
  Hertzog}, \bibinfo{person}{Sankaran~N Nair}, \bibinfo{person}{Wendy~A
  Rogers}, {and} \bibinfo{person}{Joseph Sharit}.}
  \bibinfo{year}{2006}\natexlab{}.
\newblock \showarticletitle{Factors predicting the use of technology: Findings
  from the Center for Research and Education on Aging and Technology
  Enhancement (CREATE)}.
\newblock \bibinfo{journal}{\emph{Psychology and Aging}} \bibinfo{volume}{21},
  \bibinfo{number}{2} (\bibinfo{year}{2006}), \bibinfo{pages}{333--352}.
\newblock
\urldef\tempurl%
\url{https://doi.org/10.1037/0882-7974.21.2.333}
\showDOI{\tempurl}


\bibitem[EY(2020)]%
        {EY2020}
\bibfield{author}{\bibinfo{person}{EY}.} \bibinfo{year}{2020}\natexlab{}.
\newblock \bibinfo{title}{Global FinTech Adoption Index 2019}.
\newblock
  \bibinfo{howpublished}{https://assets.ey.com/content/dam/ey-sites/ey-com/en\_gl/topics/banking-and-capital-markets/ey-global-fintech-adoption-index.pdf}.
\newblock
\newblock
\shownote{Accessed: 2023-09-07}.


\bibitem[Findlater et~al\mbox{.}(2013)]%
        {findlater2013age}
\bibfield{author}{\bibinfo{person}{Leah Findlater}, \bibinfo{person}{Jon~E
  Froehlich}, \bibinfo{person}{Kayla Fattal}, \bibinfo{person}{Jacob~O
  Wobbrock}, {and} \bibinfo{person}{Tanya Dastyar}.}
  \bibinfo{year}{2013}\natexlab{}.
\newblock \showarticletitle{Age-related differences in performance with
  touchscreens compared to traditional mouse input}. In
  \bibinfo{booktitle}{\emph{Proceedings of the SIGCHI Conference on Human
  Factors in Computing Systems}}. ACM, \bibinfo{pages}{343--346}.
\newblock


\bibitem[Franz et~al\mbox{.}(2019)]%
        {Franz2019Perception}
\bibfield{author}{\bibinfo{person}{Rachel~L. Franz}, \bibinfo{person}{Jacob~O.
  Wobbrock}, \bibinfo{person}{Yi Cheng}, {and} \bibinfo{person}{Leah
  Findlater}.} \bibinfo{year}{2019}\natexlab{}.
\newblock \showarticletitle{Perception and Adoption of Mobile Accessibility
  Features by Older Adults Experiencing Ability Changes}. In
  \bibinfo{booktitle}{\emph{Proceedings of the 21st International ACM SIGACCESS
  Conference on Computers and Accessibility}} (Pittsburgh, PA, USA)
  \emph{(\bibinfo{series}{ASSETS '19})}. \bibinfo{publisher}{Association for
  Computing Machinery}, \bibinfo{address}{New York, NY, USA},
  \bibinfo{pages}{267–278}.
\newblock
\showISBNx{9781450366762}
\urldef\tempurl%
\url{https://doi.org/10.1145/3308561.3353780}
\showDOI{\tempurl}


\bibitem[Grossi et~al\mbox{.}(2020)]%
        {grossi2020positive}
\bibfield{author}{\bibinfo{person}{Giuliano Grossi}, \bibinfo{person}{Raffaella
  Lanzarotti}, \bibinfo{person}{Paolo Napoletano}, \bibinfo{person}{Nicoletta
  Noceti}, {and} \bibinfo{person}{Francesca Odone}.}
  \bibinfo{year}{2020}\natexlab{}.
\newblock \showarticletitle{Positive technology for elderly well-being: A
  review}.
\newblock \bibinfo{journal}{\emph{Pattern recognition letters}}
  \bibinfo{volume}{137} (\bibinfo{year}{2020}), \bibinfo{pages}{61--70}.
\newblock


\bibitem[Gugenheimer et~al\mbox{.}(2022)]%
        {Gugenheimer2022Novel}
\bibfield{author}{\bibinfo{person}{Jan Gugenheimer}, \bibinfo{person}{Wen-Jie
  Tseng}, \bibinfo{person}{Abraham~Hani Mhaidli}, \bibinfo{person}{Jan~Ole
  Rixen}, \bibinfo{person}{Mark McGill}, \bibinfo{person}{Michael Nebeling},
  \bibinfo{person}{Mohamed Khamis}, \bibinfo{person}{Florian Schaub}, {and}
  \bibinfo{person}{Sanchari Das}.} \bibinfo{year}{2022}\natexlab{}.
\newblock \showarticletitle{Novel Challenges of Safety, Security and Privacy in
  Extended Reality}. In \bibinfo{booktitle}{\emph{Extended Abstracts of the
  2022 CHI Conference on Human Factors in Computing Systems}} (New Orleans, LA,
  USA) \emph{(\bibinfo{series}{CHI EA '22})}. \bibinfo{publisher}{Association
  for Computing Machinery}, \bibinfo{address}{New York, NY, USA}, Article
  \bibinfo{articleno}{108}, \bibinfo{numpages}{5}~pages.
\newblock
\showISBNx{9781450391566}
\urldef\tempurl%
\url{https://doi.org/10.1145/3491101.3503741}
\showDOI{\tempurl}


\bibitem[Guo(2017)]%
        {guoolder2017}
\bibfield{author}{\bibinfo{person}{Philip~J. Guo}.}
  \bibinfo{year}{2017}\natexlab{}.
\newblock \showarticletitle{Older {Adults} {Learning} {Computer} {Programming}:
  {Motivations}, {Frustrations}, and {Design} {Opportunities}}. In
  \bibinfo{booktitle}{\emph{Proceedings of the 2017 {CHI} {Conference} on
  {Human} {Factors} in {Computing} {Systems}}}. \bibinfo{publisher}{ACM},
  \bibinfo{address}{Denver Colorado USA}, \bibinfo{pages}{7070--7083}.
\newblock
\showISBNx{978-1-4503-4655-9}
\urldef\tempurl%
\url{https://doi.org/10.1145/3025453.3025945}
\showDOI{\tempurl}


\bibitem[Hadar(2018)]%
        {Hadar2018Toward}
\bibfield{author}{\bibinfo{person}{Ethan Hadar}.}
  \bibinfo{year}{2018}\natexlab{}.
\newblock \showarticletitle{Toward Development Tools for Augmented Reality
  Applications -- A Practitioner Perspective}. In
  \bibinfo{booktitle}{\emph{Enterprise and Organizational Modeling and
  Simulation}}, \bibfield{editor}{\bibinfo{person}{Robert Pergl},
  \bibinfo{person}{Eduard Babkin}, \bibinfo{person}{Russell Lock},
  \bibinfo{person}{Pavel Malyzhenkov}, {and} \bibinfo{person}{Vojt{\v{e}}ch
  Merunka}} (Eds.). \bibinfo{publisher}{Springer International Publishing},
  \bibinfo{address}{Cham}, \bibinfo{pages}{91--104}.
\newblock
\showISBNx{978-3-030-00787-4}


\bibitem[Han et~al\mbox{.}(2021)]%
        {Han2011Mobile}
\bibfield{author}{\bibinfo{person}{Kyungjin Han}, \bibinfo{person}{Kiho Park},
  \bibinfo{person}{Kee-Hong Choi}, {and} \bibinfo{person}{Jongweon Lee}.}
  \bibinfo{year}{2021}\natexlab{}.
\newblock \showarticletitle{Mobile Augmented Reality Serious Game for Improving
  Old Adults’ Working Memory}.
\newblock \bibinfo{journal}{\emph{Applied Sciences}} \bibinfo{volume}{11},
  \bibinfo{number}{17} (\bibinfo{year}{2021}).
\newblock
\showISSN{2076-3417}
\urldef\tempurl%
\url{https://doi.org/10.3390/app11177843}
\showDOI{\tempurl}


\bibitem[Hardy et~al\mbox{.}(2017)]%
        {Hardy2017Baby}
\bibfield{author}{\bibinfo{person}{M. Hardy}, \bibinfo{person}{F. Oprescu},
  \bibinfo{person}{P. Millear}, {and} \bibinfo{person}{M. Summers}.}
  \bibinfo{year}{2017}\natexlab{}.
\newblock \showarticletitle{Baby boomers engagement as traditional university
  students: Benefits and costs}.
\newblock \bibinfo{journal}{\emph{International Journal of Lifelong Education}}
  \bibinfo{volume}{36}, \bibinfo{number}{6} (\bibinfo{year}{2017}),
  \bibinfo{pages}{730--744}.
\newblock
\urldef\tempurl%
\url{https://doi.org/10.1080/02601370.2017.1382015}
\showDOI{\tempurl}


\bibitem[Hartmann et~al\mbox{.}(2020)]%
        {hartmann2020extend}
\bibfield{author}{\bibinfo{person}{Jeremy Hartmann}, \bibinfo{person}{Aakar
  Gupta}, {and} \bibinfo{person}{Daniel Vogel}.}
  \bibinfo{year}{2020}\natexlab{}.
\newblock \showarticletitle{Extend, push, pull: smartphone mediated interaction
  in spatial augmented reality via intuitive mode switching}. In
  \bibinfo{booktitle}{\emph{Proceedings of the 2020 ACM Symposium on Spatial
  User Interaction}}. \bibinfo{pages}{1--10}.
\newblock


\bibitem[He et~al\mbox{.}(2023)]%
        {he2023ubi}
\bibfield{author}{\bibinfo{person}{Fengming He}, \bibinfo{person}{Xiyun Hu},
  \bibinfo{person}{Jingyu Shi}, \bibinfo{person}{Xun Qian},
  \bibinfo{person}{Tianyi Wang}, {and} \bibinfo{person}{Karthik Ramani}.}
  \bibinfo{year}{2023}\natexlab{}.
\newblock \showarticletitle{Ubi Edge: Authoring Edge-Based Opportunistic
  Tangible User Interfaces in Augmented Reality}. In
  \bibinfo{booktitle}{\emph{Proceedings of the 2023 CHI Conference on Human
  Factors in Computing Systems}}. \bibinfo{pages}{1--14}.
\newblock


\bibitem[Hollinworth and Hwang(2010)]%
        {hollinworthrelating2010}
\bibfield{author}{\bibinfo{person}{*~Nic Hollinworth} {and}
  \bibinfo{person}{Faustina Hwang}.} \bibinfo{year}{2010}\natexlab{}.
\newblock \showarticletitle{Relating computer tasks to existing knowledge to
  improve accessibility for older adults}. In
  \bibinfo{booktitle}{\emph{Proceedings of the 12th international {ACM}
  {SIGACCESS} conference on {Computers} and accessibility}}.
  \bibinfo{publisher}{ACM}, \bibinfo{address}{Orlando Florida USA},
  \bibinfo{pages}{147--154}.
\newblock
\showISBNx{978-1-60558-881-0}
\urldef\tempurl%
\url{https://doi.org/10.1145/1878803.1878830}
\showDOI{\tempurl}


\bibitem[Huang et~al\mbox{.}(2021)]%
        {huang2021adaptutar}
\bibfield{author}{\bibinfo{person}{Gaoping Huang}, \bibinfo{person}{Xun Qian},
  \bibinfo{person}{Tianyi Wang}, \bibinfo{person}{Fagun Patel},
  \bibinfo{person}{Maitreya Sreeram}, \bibinfo{person}{Yuanzhi Cao},
  \bibinfo{person}{Karthik Ramani}, {and} \bibinfo{person}{Alexander~J Quinn}.}
  \bibinfo{year}{2021}\natexlab{}.
\newblock \showarticletitle{Adaptutar: An adaptive tutoring system for machine
  tasks in augmented reality}. In \bibinfo{booktitle}{\emph{Proceedings of the
  2021 CHI Conference on Human Factors in Computing Systems}}.
  \bibinfo{pages}{1--15}.
\newblock


\bibitem[Hubenschmid et~al\mbox{.}(2023)]%
        {hubenschmid2023around}
\bibfield{author}{\bibinfo{person}{Sebastian Hubenschmid},
  \bibinfo{person}{Johannes Zagermann}, \bibinfo{person}{Daniel Leicht},
  \bibinfo{person}{Harald Reiterer}, {and} \bibinfo{person}{Tiare Feuchtner}.}
  \bibinfo{year}{2023}\natexlab{}.
\newblock \showarticletitle{ARound the Smartphone: Investigating the Effects of
  Virtually-Extended Display Size on Spatial Memory}. In
  \bibinfo{booktitle}{\emph{Proceedings of the 2023 CHI Conference on Human
  Factors in Computing Systems}}. \bibinfo{pages}{1--15}.
\newblock


\bibitem[Irawati et~al\mbox{.}(2006)]%
        {irawati2006evaluation}
\bibfield{author}{\bibinfo{person}{Sylvia Irawati}, \bibinfo{person}{Scott
  Green}, \bibinfo{person}{Mark Billinghurst}, \bibinfo{person}{Andreas
  Duenser}, {and} \bibinfo{person}{Heedong Ko}.}
  \bibinfo{year}{2006}\natexlab{}.
\newblock \showarticletitle{An evaluation of an augmented reality multimodal
  interface using speech and paddle gestures}. In
  \bibinfo{booktitle}{\emph{Advances in Artificial Reality and Tele-Existence:
  16th International Conference on Artificial Reality and Telexistence, ICAT
  2006, Hangzhou, China, November 29-December 1, 2006. Proceedings}}. Springer,
  \bibinfo{pages}{272--283}.
\newblock


\bibitem[Jin and Fan(2022)]%
        {Jin22I}
\bibfield{author}{\bibinfo{person}{Xiaofu Jin} {and} \bibinfo{person}{Mingming
  Fan}.} \bibinfo{year}{2022}\natexlab{}.
\newblock \showarticletitle{“I Used To Carry A Wallet, Now I Just Need To
  Carry My Phone”: Understanding Current Banking Practices and Challenges
  Among Older Adults in China}. In \bibinfo{booktitle}{\emph{Proceedings of the
  24th International ACM SIGACCESS Conference on Computers and Accessibility}}
  (Athens, Greece) \emph{(\bibinfo{series}{ASSETS '22})}.
  \bibinfo{publisher}{Association for Computing Machinery},
  \bibinfo{address}{New York, NY, USA}, Article \bibinfo{articleno}{37},
  \bibinfo{numpages}{16}~pages.
\newblock
\showISBNx{9781450392587}
\urldef\tempurl%
\url{https://doi.org/10.1145/3517428.3544820}
\showDOI{\tempurl}


\bibitem[Jin et~al\mbox{.}(2022)]%
        {jinsynapse2022}
\bibfield{author}{\bibinfo{person}{Xiaofu Jin}, \bibinfo{person}{Xiaozhu Hu},
  \bibinfo{person}{Xiaoying Wei}, {and} \bibinfo{person}{Mingming Fan}.}
  \bibinfo{year}{2022}\natexlab{}.
\newblock \showarticletitle{Synapse: {Interactive} {Guidance} by
  {Demonstration} with {Trial}-and-{Error} {Support} for {Older} {Adults} to
  {Use} {Smartphone} {Apps}}.
\newblock \bibinfo{journal}{\emph{Proceedings of the ACM on Interactive,
  Mobile, Wearable and Ubiquitous Technologies}} \bibinfo{volume}{6},
  \bibinfo{number}{3} (\bibinfo{date}{Sept.} \bibinfo{year}{2022}),
  \bibinfo{pages}{1--24}.
\newblock
\showISSN{2474-9567}
\urldef\tempurl%
\url{https://doi.org/10.1145/3550321}
\showDOI{\tempurl}


\bibitem[Kane et~al\mbox{.}(2008)]%
        {kane2008slide}
\bibfield{author}{\bibinfo{person}{Shaun~K Kane}, \bibinfo{person}{Jeffrey~P
  Bigham}, {and} \bibinfo{person}{Jacob~O Wobbrock}.}
  \bibinfo{year}{2008}\natexlab{}.
\newblock \showarticletitle{Slide rule: making mobile touch screens accessible
  to blind people using multi-touch interaction techniques}. In
  \bibinfo{booktitle}{\emph{Proceedings of the 10th international ACM SIGACCESS
  conference on Computers and accessibility}}. \bibinfo{pages}{73--80}.
\newblock


\bibitem[Khan et~al\mbox{.}(2018)]%
        {Khan2018Mathland}
\bibfield{author}{\bibinfo{person}{Mina Khan}, \bibinfo{person}{Fernando
  Trujano}, \bibinfo{person}{Ashris Choudhury}, {and} \bibinfo{person}{Pattie
  Maes}.} \bibinfo{year}{2018}\natexlab{}.
\newblock \showarticletitle{Mathland: Playful Mathematical Learning in Mixed
  Reality}. In \bibinfo{booktitle}{\emph{Extended Abstracts of the 2018 CHI
  Conference on Human Factors in Computing Systems}} (Montreal QC, Canada)
  \emph{(\bibinfo{series}{CHI EA '18})}. \bibinfo{publisher}{Association for
  Computing Machinery}, \bibinfo{address}{New York, NY, USA},
  \bibinfo{pages}{1–4}.
\newblock
\showISBNx{9781450356213}
\urldef\tempurl%
\url{https://doi.org/10.1145/3170427.3186499}
\showDOI{\tempurl}


\bibitem[Klopfer and Squire(2008)]%
        {klopfer2008augmented}
\bibfield{author}{\bibinfo{person}{Eric Klopfer} {and} \bibinfo{person}{Kurt
  Squire}.} \bibinfo{year}{2008}\natexlab{}.
\newblock \showarticletitle{Augmented reality: Research and design of mobile
  educational games}.
\newblock \bibinfo{journal}{\emph{The TechTrends}} \bibinfo{volume}{49},
  \bibinfo{number}{3} (\bibinfo{year}{2008}), \bibinfo{pages}{38--45}.
\newblock


\bibitem[Kobayashi et~al\mbox{.}(2011)]%
        {kobayashielderly2011}
\bibfield{author}{\bibinfo{person}{Masatomo Kobayashi},
  \bibinfo{person}{Atsushi Hiyama}, \bibinfo{person}{Takahiro Miura},
  \bibinfo{person}{Chieko Asakawa}, \bibinfo{person}{Michitaka Hirose}, {and}
  \bibinfo{person}{Tohru Ifukube}.} \bibinfo{year}{2011}\natexlab{}.
\newblock \showarticletitle{Elderly {User} {Evaluation} of {Mobile}
  {Touchscreen} {Interactions}}. In \bibinfo{booktitle}{\emph{Human-{Computer}
  {Interaction} – {INTERACT} 2011}} \emph{(\bibinfo{series}{Lecture {Notes}
  in {Computer} {Science}})}, \bibfield{editor}{\bibinfo{person}{Pedro Campos},
  \bibinfo{person}{Nicholas Graham}, \bibinfo{person}{Joaquim Jorge},
  \bibinfo{person}{Nuno Nunes}, \bibinfo{person}{Philippe Palanque}, {and}
  \bibinfo{person}{Marco Winckler}} (Eds.). \bibinfo{publisher}{Springer},
  \bibinfo{address}{Berlin, Heidelberg}, \bibinfo{pages}{83--99}.
\newblock
\showISBNx{978-3-642-23774-4}
\urldef\tempurl%
\url{https://doi.org/10.1007/978-3-642-23774-4\_9}
\showDOI{\tempurl}


\bibitem[Kopeć et~al\mbox{.}(2017)]%
        {kopeclivinglab2017}
\bibfield{author}{\bibinfo{person}{Wiesław Kopeć}, \bibinfo{person}{Kinga
  Skorupska}, \bibinfo{person}{Anna Jaskulska}, \bibinfo{person}{Katarzyna
  Abramczuk}, \bibinfo{person}{Radoslaw Nielek}, {and} \bibinfo{person}{Adam
  Wierzbicki}.} \bibinfo{year}{2017}\natexlab{}.
\newblock \showarticletitle{{LivingLab} {PJAIT}: towards better urban
  participation of seniors}. In \bibinfo{booktitle}{\emph{Proceedings of the
  {International} {Conference} on {Web} {Intelligence}}}.
  \bibinfo{publisher}{ACM}, \bibinfo{address}{Leipzig Germany},
  \bibinfo{pages}{1085--1092}.
\newblock
\showISBNx{978-1-4503-4951-2}
\urldef\tempurl%
\url{https://doi.org/10.1145/3106426.3109040}
\showDOI{\tempurl}


\bibitem[K{\"u}{\c{c}}{\"u}k et~al\mbox{.}(2016)]%
        {kucuk2016educational}
\bibfield{author}{\bibinfo{person}{Serdar K{\"u}{\c{c}}{\"u}k},
  \bibinfo{person}{Samet Kapakin}, {and} \bibinfo{person}{Y{\"u}ksel
  G{\"o}ktas}.} \bibinfo{year}{2016}\natexlab{}.
\newblock \showarticletitle{Educational Augmentation of Anatomy Learning: An
  Analysis with Augmented Reality Techniques}.
\newblock \bibinfo{journal}{\emph{Anatolian Journal of Cardiology}}
  \bibinfo{volume}{16}, \bibinfo{number}{4} (\bibinfo{year}{2016}),
  \bibinfo{pages}{123}.
\newblock


\bibitem[Lee et~al\mbox{.}(2019)]%
        {Lee2019Potential}
\bibfield{author}{\bibinfo{person}{Li~Na Lee}, \bibinfo{person}{Mi~Jeong Kim},
  {and} \bibinfo{person}{Won~Ju Hwang}.} \bibinfo{year}{2019}\natexlab{}.
\newblock \showarticletitle{Potential of Augmented Reality and Virtual Reality
  Technologies to Promote Wellbeing in Older Adults}.
\newblock \bibinfo{journal}{\emph{Applied Sciences}} \bibinfo{volume}{9},
  \bibinfo{number}{17} (\bibinfo{year}{2019}).
\newblock
\showISSN{2076-3417}
\urldef\tempurl%
\url{https://doi.org/10.3390/app9173556}
\showDOI{\tempurl}


\bibitem[Lee and Billinghurst(2008)]%
        {lee2008wizard}
\bibfield{author}{\bibinfo{person}{Minkyung Lee} {and} \bibinfo{person}{Mark
  Billinghurst}.} \bibinfo{year}{2008}\natexlab{}.
\newblock \showarticletitle{A wizard of oz study for an ar multimodal
  interface}. In \bibinfo{booktitle}{\emph{Proceedings of the 10th
  international conference on Multimodal interfaces}}.
  \bibinfo{pages}{249--256}.
\newblock


\bibitem[Leitão and Silva(2012)]%
        {targetleito2012}
\bibfield{author}{\bibinfo{person}{Roxanne Leitão} {and}
  \bibinfo{person}{Paula~Alexandra Silva}.} \bibinfo{year}{2012}\natexlab{}.
\newblock \showarticletitle{Target and spacing sizes for smartphone user
  interfaces for older adults: design patterns based on an evaluation with
  users}.
\newblock  (\bibinfo{year}{2012}).
\newblock
\urldef\tempurl%
\url{https://doi.org/10.5555/2821679.2831275}
\showDOI{\tempurl}


\bibitem[Leung et~al\mbox{.}(2012)]%
        {leunghow2012}
\bibfield{author}{\bibinfo{person}{Rock Leung}, \bibinfo{person}{Charlotte
  Tang}, \bibinfo{person}{Shathel Haddad}, \bibinfo{person}{Joanna Mcgrenere},
  \bibinfo{person}{Peter Graf}, {and} \bibinfo{person}{Vilia Ingriany}.}
  \bibinfo{year}{2012}\natexlab{}.
\newblock \showarticletitle{How {Older} {Adults} {Learn} to {Use} {Mobile}
  {Devices}: {Survey} and {Field} {Investigations}}.
\newblock \bibinfo{journal}{\emph{ACM Transactions on Accessible Computing}}
  \bibinfo{volume}{4}, \bibinfo{number}{3} (\bibinfo{date}{Dec.}
  \bibinfo{year}{2012}), \bibinfo{pages}{1--33}.
\newblock
\showISSN{1936-7228, 1936-7236}
\urldef\tempurl%
\url{https://doi.org/10.1145/2399193.2399195}
\showDOI{\tempurl}


\bibitem[Lin et~al\mbox{.}(2020)]%
        {Lin2020Exploring}
\bibfield{author}{\bibinfo{person}{Kuan-Yu Lin}, \bibinfo{person}{Yi-Ting
  Wang}, {and} \bibinfo{person}{Travis~K. Huang}.}
  \bibinfo{year}{2020}\natexlab{}.
\newblock \showarticletitle{Exploring the antecedents of mobile payment service
  usage: Perspectives based on cost–benefit theory, perceived value, and
  social influences}.
\newblock \bibinfo{journal}{\emph{Online information review}}
  \bibinfo{volume}{44}, \bibinfo{number}{1} (\bibinfo{year}{2020}),
  \bibinfo{pages}{299--318}.
\newblock
\showISBNx{1468-4527}


\bibitem[Lin et~al\mbox{.}(2018)]%
        {linwhy2018}
\bibfield{author}{\bibinfo{person}{Yu-Hao Lin}, \bibinfo{person}{Suwen Zhu},
  \bibinfo{person}{Yu-Jung Ko}, \bibinfo{person}{Wenzhe Cui}, {and}
  \bibinfo{person}{Xiaojun Bi}.} \bibinfo{year}{2018}\natexlab{}.
\newblock \showarticletitle{Why {Is} {Gesture} {Typing} {Promising} for {Older}
  {Adults}?: {Comparing} {Gesture} and {Tap} {Typing} {Behavior} of {Older}
  with {Young} {Adults}}. In \bibinfo{booktitle}{\emph{Proceedings of the 20th
  {International} {ACM} {SIGACCESS} {Conference} on {Computers} and
  {Accessibility}}}. \bibinfo{publisher}{ACM}, \bibinfo{address}{Galway
  Ireland}, \bibinfo{pages}{271--281}.
\newblock
\showISBNx{978-1-4503-5650-3}
\urldef\tempurl%
\url{https://doi.org/10.1145/3234695.3236350}
\showDOI{\tempurl}


\bibitem[Liu et~al\mbox{.}(2023)]%
        {liu2023instrumentar}
\bibfield{author}{\bibinfo{person}{Ziyi Liu}, \bibinfo{person}{Zhengzhe Zhu},
  \bibinfo{person}{Enze Jiang}, \bibinfo{person}{Feichi Huang},
  \bibinfo{person}{Ana~M Villanueva}, \bibinfo{person}{Xun Qian},
  \bibinfo{person}{Tianyi Wang}, {and} \bibinfo{person}{Karthik Ramani}.}
  \bibinfo{year}{2023}\natexlab{}.
\newblock \showarticletitle{InstruMentAR: Auto-Generation of Augmented Reality
  Tutorials for Operating Digital Instruments Through Recording Embodied
  Demonstration}. In \bibinfo{booktitle}{\emph{Proceedings of the 2023 CHI
  Conference on Human Factors in Computing Systems}}. \bibinfo{pages}{1--17}.
\newblock


\bibitem[Mahmud et~al\mbox{.}(2020)]%
        {mahmud2020learning}
\bibfield{author}{\bibinfo{person}{Shareen Mahmud}, \bibinfo{person}{Jessalyn
  Alvina}, \bibinfo{person}{Parmit~K Chilana}, \bibinfo{person}{Andrea Bunt},
  {and} \bibinfo{person}{Joanna McGrenere}.} \bibinfo{year}{2020}\natexlab{}.
\newblock \showarticletitle{Learning through exploration: how children, adults,
  and older adults interact with a new feature-rich application}. In
  \bibinfo{booktitle}{\emph{Proceedings of the 2020 CHI Conference on Human
  Factors in Computing Systems}}. \bibinfo{pages}{1--14}.
\newblock


\bibitem[Mendel and Toch(2022)]%
        {mendelmeerkat2022}
\bibfield{author}{\bibinfo{person}{Tamir Mendel} {and} \bibinfo{person}{Eran
  Toch}.} \bibinfo{year}{2022}\natexlab{}.
\newblock \showarticletitle{Meerkat: {A} {Social} {Community} {Support}
  {Application} for {Older} {Adults}}. In \bibinfo{booktitle}{\emph{{CHI}
  {Conference} on {Human} {Factors} in {Computing} {Systems} {Extended}
  {Abstracts}}}. \bibinfo{publisher}{ACM}, \bibinfo{address}{New Orleans LA
  USA}, \bibinfo{pages}{1--4}.
\newblock
\showISBNx{978-1-4503-9156-6}
\urldef\tempurl%
\url{https://doi.org/10.1145/3491101.3519909}
\showDOI{\tempurl}


\bibitem[Menghi et~al\mbox{.}(2017)]%
        {menghihow2017}
\bibfield{author}{\bibinfo{person}{Roberto Menghi}, \bibinfo{person}{Silvia
  Ceccacci}, \bibinfo{person}{Francesca Gullà}, \bibinfo{person}{Lorenzo
  Cavalieri}, \bibinfo{person}{Michele Germani}, {and} \bibinfo{person}{Roberta
  Bevilacqua}.} \bibinfo{year}{2017}\natexlab{}.
\newblock \showarticletitle{How {Older} {People} {Who} {Have} {Never} {Used}
  {Touchscreen} {Technology} {Interact} with a {Tablet}}. In
  \bibinfo{booktitle}{\emph{Human-{Computer} {Interaction} - {INTERACT} 2017}}
  \emph{(\bibinfo{series}{Lecture {Notes} in {Computer} {Science}})},
  \bibfield{editor}{\bibinfo{person}{Regina Bernhaupt}, \bibinfo{person}{Girish
  Dalvi}, \bibinfo{person}{Anirudha Joshi}, \bibinfo{person}{Devanuj
  K.~Balkrishan}, \bibinfo{person}{Jacki O'Neill}, {and} \bibinfo{person}{Marco
  Winckler}} (Eds.). \bibinfo{publisher}{Springer International Publishing},
  \bibinfo{address}{Cham}, \bibinfo{pages}{117--131}.
\newblock
\showISBNx{978-3-319-67744-6}
\urldef\tempurl%
\url{https://doi.org/10.1007/978-3-319-67744-68}
\showDOI{\tempurl}


\bibitem[Microsoft(2019)]%
        {microsoft2019hololens}
\bibfield{author}{\bibinfo{person}{Microsoft}.}
  \bibinfo{year}{2019}\natexlab{}.
\newblock \bibinfo{booktitle}{\emph{Getting started with HoloLens 2}}.
\newblock
\urldef\tempurl%
\url{https://www.microsoft.com/en-us/hololens/}
\showURL{%
\tempurl}
\newblock
\shownote{Accessed: 2023-9-7}.


\bibitem[Mitzner et~al\mbox{.}(2010)]%
        {mitzner2010older}
\bibfield{author}{\bibinfo{person}{Tracy~L Mitzner}, \bibinfo{person}{Julie~B
  Boron}, \bibinfo{person}{Cara~B Fausset}, \bibinfo{person}{Anne~E Adams},
  \bibinfo{person}{Neil Charness}, \bibinfo{person}{Sara~J Czaja}, {and}
  \bibinfo{person}{Wendy~A Rogers}.} \bibinfo{year}{2010}\natexlab{}.
\newblock \showarticletitle{Older adults talk technology: Technology usage and
  attitudes}.
\newblock \bibinfo{journal}{\emph{Computers in Human Behavior}}
  \bibinfo{volume}{26}, \bibinfo{number}{6} (\bibinfo{year}{2010}),
  \bibinfo{pages}{1710--1721}.
\newblock


\bibitem[Mitzner et~al\mbox{.}(2018)]%
        {Mitzner2018Technology}
\bibfield{author}{\bibinfo{person}{Tracy~L Mitzner}, \bibinfo{person}{Jyoti
  Savla}, \bibinfo{person}{Walter~R Boot}, \bibinfo{person}{Joseph Sharit},
  \bibinfo{person}{Neil Charness}, \bibinfo{person}{Sara~J Czaja}, {and}
  \bibinfo{person}{Wendy~A Rogers}.} \bibinfo{year}{2018}\natexlab{}.
\newblock \showarticletitle{{Technology Adoption by Older Adults: Findings From
  the PRISM Trial}}.
\newblock \bibinfo{journal}{\emph{The Gerontologist}} \bibinfo{volume}{59},
  \bibinfo{number}{1} (\bibinfo{date}{09} \bibinfo{year}{2018}),
  \bibinfo{pages}{34--44}.
\newblock
\showISSN{0016-9013}
\urldef\tempurl%
\url{https://doi.org/10.1093/geront/gny113}
\showDOI{\tempurl}
\showeprint{https://academic.oup.com/gerontologist/article-pdf/59/1/34/27456596/gny113.pdf}


\bibitem[Mostajeran et~al\mbox{.}(2020)]%
        {Mostajeran2020Augmented}
\bibfield{author}{\bibinfo{person}{Fariba Mostajeran}, \bibinfo{person}{Frank
  Steinicke}, \bibinfo{person}{Oscar~Javier Ariza~Nunez},
  \bibinfo{person}{Dimitrios Gatsios}, {and} \bibinfo{person}{Dimitrios
  Fotiadis}.} \bibinfo{year}{2020}\natexlab{}.
\newblock \showarticletitle{Augmented Reality for Older Adults: Exploring
  Acceptability of Virtual Coaches for Home-Based Balance Training in an Aging
  Population}. In \bibinfo{booktitle}{\emph{Proceedings of the 2020 CHI
  Conference on Human Factors in Computing Systems}} (Honolulu, HI, USA)
  \emph{(\bibinfo{series}{CHI '20})}. \bibinfo{publisher}{Association for
  Computing Machinery}, \bibinfo{address}{New York, NY, USA},
  \bibinfo{pages}{1–12}.
\newblock
\showISBNx{9781450367080}
\urldef\tempurl%
\url{https://doi.org/10.1145/3313831.3376565}
\showDOI{\tempurl}


\bibitem[Müller et~al\mbox{.}(2015)]%
        {mullerpractice-based2015}
\bibfield{author}{\bibinfo{person}{Claudia Müller}, \bibinfo{person}{Dominik
  Hornung}, \bibinfo{person}{Theodor Hamm}, {and} \bibinfo{person}{Volker
  Wulf}.} \bibinfo{year}{2015}\natexlab{}.
\newblock \showarticletitle{Practice-based {Design} of a {Neighborhood}
  {Portal}: {Focusing} on {Elderly} {Tenants} in a {City} {Quarter} {Living}
  {Lab}}. In \bibinfo{booktitle}{\emph{Proceedings of the 33rd {Annual} {ACM}
  {Conference} on {Human} {Factors} in {Computing} {Systems}}}.
  \bibinfo{publisher}{ACM}, \bibinfo{address}{Seoul Republic of Korea},
  \bibinfo{pages}{2295--2304}.
\newblock
\showISBNx{978-1-4503-3145-6}
\urldef\tempurl%
\url{https://doi.org/10.1145/2702123.2702449}
\showDOI{\tempurl}


\bibitem[Nishimoto and Johnson(2019)]%
        {nishimoto2019extending}
\bibfield{author}{\bibinfo{person}{Arthur Nishimoto} {and}
  \bibinfo{person}{Andrew~E Johnson}.} \bibinfo{year}{2019}\natexlab{}.
\newblock \showarticletitle{Extending virtual reality display wall environments
  using augmented reality}. In \bibinfo{booktitle}{\emph{Symposium on spatial
  user interaction}}. \bibinfo{pages}{1--5}.
\newblock


\bibitem[Norman(2010)]%
        {norman2010natural}
\bibfield{author}{\bibinfo{person}{Donald~A Norman}.}
  \bibinfo{year}{2010}\natexlab{}.
\newblock \showarticletitle{Natural user interfaces are not natural}.
\newblock \bibinfo{journal}{\emph{interactions}} \bibinfo{volume}{17},
  \bibinfo{number}{3} (\bibinfo{year}{2010}), \bibinfo{pages}{6--10}.
\newblock


\bibitem[O'Brien and Burn(2012)]%
        {o'brien2012}
\bibfield{author}{\bibinfo{person}{Ames~D. O'Brien, J.~G.} {and}
  \bibinfo{person}{A. Burn}.} \bibinfo{year}{2012}\natexlab{}.
\newblock \showarticletitle{Clinical Management of Age-Related Macular
  Degeneration}.
\newblock \bibinfo{journal}{\emph{The Medical Journal of Australia}}
  \bibinfo{volume}{196} (\bibinfo{year}{2012}), \bibinfo{pages}{322--324}.
\newblock


\bibitem[Ordonez et~al\mbox{.}(2011)]%
        {ordonez2011elderly}
\bibfield{author}{\bibinfo{person}{Tiago~Nascimento Ordonez},
  \bibinfo{person}{M{\^o}nica~Sanches Yassuda}, {and} \bibinfo{person}{Meire
  Cachioni}.} \bibinfo{year}{2011}\natexlab{}.
\newblock \showarticletitle{Elderly online: effects of a digital inclusion
  program in cognitive performance}.
\newblock \bibinfo{journal}{\emph{Archives of Gerontology and Geriatrics}}
  \bibinfo{volume}{53}, \bibinfo{number}{2} (\bibinfo{year}{2011}),
  \bibinfo{pages}{216--219}.
\newblock


\bibitem[O’Brien et~al\mbox{.}(2008)]%
        {o2008understanding}
\bibfield{author}{\bibinfo{person}{Marita~A O’Brien},
  \bibinfo{person}{Katherine~E Olson}, \bibinfo{person}{Neil Charness},
  \bibinfo{person}{Sara~J Czaja}, \bibinfo{person}{Arthur~D Fisk},
  \bibinfo{person}{Wendy~A Rogers}, {and} \bibinfo{person}{Joseph Sharit}.}
  \bibinfo{year}{2008}\natexlab{}.
\newblock \showarticletitle{Understanding technology usage in older adults}.
\newblock \bibinfo{journal}{\emph{Proceedings of the 6th International Society
  for Gerontechnology, Pisa, Italy}} (\bibinfo{year}{2008}).
\newblock


\bibitem[Palumbo and Paterno(2021)]%
        {palumbomicogito2021}
\bibfield{author}{\bibinfo{person}{*~Vanessa Palumbo} {and}
  \bibinfo{person}{Fabio Paterno}.} \bibinfo{year}{2021}\natexlab{}.
\newblock \showarticletitle{Micogito: a {Serious} {Gamebook} {Based} on {Daily}
  {Life} {Scenarios} to {Cognitively} {Stimulate} {Older} {Adults}}. In
  \bibinfo{booktitle}{\emph{Proceedings of the {Conference} on {Information}
  {Technology} for {Social} {Good}}}. \bibinfo{publisher}{ACM},
  \bibinfo{address}{Roma Italy}, \bibinfo{pages}{163--168}.
\newblock
\showISBNx{978-1-4503-8478-0}
\urldef\tempurl%
\url{https://doi.org/10.1145/3462203.3475889}
\showDOI{\tempurl}


\bibitem[Pandya and El-Glaly(2018)]%
        {pandyataptag2018}
\bibfield{author}{\bibinfo{person}{Shraddha Pandya} {and}
  \bibinfo{person}{Yasmine~N. El-Glaly}.} \bibinfo{year}{2018}\natexlab{}.
\newblock \showarticletitle{TapTag: Assistive Gestural Interactions in Social
  Media on Touchscreens for Older Adults}. In
  \bibinfo{booktitle}{\emph{Proceedings of the 20th {ACM} {International}
  {Conference} on {Multimodal} {Interaction}}}. \bibinfo{publisher}{ACM},
  \bibinfo{address}{Boulder CO USA}, \bibinfo{pages}{244--252}.
\newblock
\showISBNx{978-1-4503-5692-3}
\urldef\tempurl%
\url{https://doi.org/10.1145/3242969.3243003}
\showDOI{\tempurl}


\bibitem[Pang et~al\mbox{.}(2021)]%
        {Pang2021Technology}
\bibfield{author}{\bibinfo{person}{Carolyn Pang}, \bibinfo{person}{Zhiqin
  Collin~Wang}, \bibinfo{person}{Joanna McGrenere}, \bibinfo{person}{Rock
  Leung}, \bibinfo{person}{Jiamin Dai}, {and} \bibinfo{person}{Karyn Moffatt}.}
  \bibinfo{year}{2021}\natexlab{}.
\newblock \showarticletitle{Technology Adoption and Learning Preferences for
  Older Adults: Evolving Perceptions, Ongoing Challenges, and Emerging Design
  Opportunities}. In \bibinfo{booktitle}{\emph{Proceedings of the 2021 CHI
  Conference on Human Factors in Computing Systems}} (Yokohama, Japan)
  \emph{(\bibinfo{series}{CHI '21})}. \bibinfo{publisher}{Association for
  Computing Machinery}, \bibinfo{address}{New York, NY, USA}, Article
  \bibinfo{articleno}{490}, \bibinfo{numpages}{13}~pages.
\newblock
\showISBNx{9781450380966}
\urldef\tempurl%
\url{https://doi.org/10.1145/3411764.3445702}
\showDOI{\tempurl}


\bibitem[Piumsomboon et~al\mbox{.}(2014)]%
        {piumsomboon2014grasp}
\bibfield{author}{\bibinfo{person}{Thammathip Piumsomboon},
  \bibinfo{person}{David Altimira}, \bibinfo{person}{Hyungon Kim},
  \bibinfo{person}{Adrian Clark}, \bibinfo{person}{Gun Lee}, {and}
  \bibinfo{person}{Mark Billinghurst}.} \bibinfo{year}{2014}\natexlab{}.
\newblock \showarticletitle{Grasp-Shell vs gesture-speech: A comparison of
  direct and indirect natural interaction techniques in augmented reality}. In
  \bibinfo{booktitle}{\emph{2014 IEEE International Symposium on Mixed and
  Augmented Reality (ISMAR)}}. IEEE, \bibinfo{pages}{73--82}.
\newblock


\bibitem[Pradhan et~al\mbox{.}(2020)]%
        {Pradhan2020Understanding}
\bibfield{author}{\bibinfo{person}{Alisha Pradhan}, \bibinfo{person}{Ben
  Jelen}, \bibinfo{person}{Katie~A. Siek}, \bibinfo{person}{Joel Chan}, {and}
  \bibinfo{person}{Amanda Lazar}.} \bibinfo{year}{2020}\natexlab{}.
\newblock \showarticletitle{Understanding Older Adults' Participation in Design
  Workshops}. In \bibinfo{booktitle}{\emph{Proceedings of the 2020 CHI
  Conference on Human Factors in Computing Systems}} (Honolulu, HI, USA)
  \emph{(\bibinfo{series}{CHI '20})}. \bibinfo{publisher}{Association for
  Computing Machinery}, \bibinfo{address}{New York, NY, USA},
  \bibinfo{pages}{1–15}.
\newblock
\showISBNx{9781450367080}
\urldef\tempurl%
\url{https://doi.org/10.1145/3313831.3376299}
\showDOI{\tempurl}


\bibitem[Radu and Schneider(2019)]%
        {Radu2019What}
\bibfield{author}{\bibinfo{person}{Iulian Radu} {and} \bibinfo{person}{Bertrand
  Schneider}.} \bibinfo{year}{2019}\natexlab{}.
\newblock \showarticletitle{What Can We Learn from Augmented Reality (AR)?
  Benefits and Drawbacks of AR for Inquiry-Based Learning of Physics}. In
  \bibinfo{booktitle}{\emph{Proceedings of the 2019 CHI Conference on Human
  Factors in Computing Systems}} (Glasgow, Scotland Uk)
  \emph{(\bibinfo{series}{CHI '19})}. \bibinfo{publisher}{Association for
  Computing Machinery}, \bibinfo{address}{New York, NY, USA},
  \bibinfo{pages}{1–12}.
\newblock
\showISBNx{9781450359702}
\urldef\tempurl%
\url{https://doi.org/10.1145/3290605.3300774}
\showDOI{\tempurl}


\bibitem[Rajaram et~al\mbox{.}(2023)]%
        {Rajaram2023Eliciting}
\bibfield{author}{\bibinfo{person}{Shwetha Rajaram}, \bibinfo{person}{Chen
  Chen}, \bibinfo{person}{Franziska Roesner}, {and} \bibinfo{person}{Michael
  Nebeling}.} \bibinfo{year}{2023}\natexlab{}.
\newblock \showarticletitle{Eliciting Security \& Privacy-Informed Sharing
  Techniques for Multi-User Augmented Reality}. In
  \bibinfo{booktitle}{\emph{Proceedings of the 2023 CHI Conference on Human
  Factors in Computing Systems}} (Hamburg, Germany) \emph{(\bibinfo{series}{CHI
  '23})}. \bibinfo{publisher}{Association for Computing Machinery},
  \bibinfo{address}{New York, NY, USA}, Article \bibinfo{articleno}{98},
  \bibinfo{numpages}{17}~pages.
\newblock
\showISBNx{9781450394215}
\urldef\tempurl%
\url{https://doi.org/10.1145/3544548.3581089}
\showDOI{\tempurl}


\bibitem[Reichherzer et~al\mbox{.}(2021)]%
        {reichherzer2021secondsight}
\bibfield{author}{\bibinfo{person}{Carolin Reichherzer}, \bibinfo{person}{Jack
  Fraser}, \bibinfo{person}{Damien~Constantine Rompapas}, {and}
  \bibinfo{person}{Mark Billinghurst}.} \bibinfo{year}{2021}\natexlab{}.
\newblock \showarticletitle{Secondsight: A framework for cross-device augmented
  reality interfaces}. In \bibinfo{booktitle}{\emph{Extended Abstracts of the
  2021 CHI Conference on Human Factors in Computing Systems}}.
  \bibinfo{pages}{1--6}.
\newblock


\bibitem[Reipschl{\"a}ger and Dachselt(2019)]%
        {reipschlager2019designar}
\bibfield{author}{\bibinfo{person}{Patrick Reipschl{\"a}ger} {and}
  \bibinfo{person}{Raimund Dachselt}.} \bibinfo{year}{2019}\natexlab{}.
\newblock \showarticletitle{Designar: Immersive 3d-modeling combining augmented
  reality with interactive displays}. In \bibinfo{booktitle}{\emph{Proceedings
  of the 2019 ACM International Conference on Interactive Surfaces and
  Spaces}}. \bibinfo{pages}{29--41}.
\newblock


\bibitem[Reipschl{\"a}ger et~al\mbox{.}(2020)]%
        {reipschlager2020augmented}
\bibfield{author}{\bibinfo{person}{Patrick Reipschl{\"a}ger},
  \bibinfo{person}{Severin Engert}, {and} \bibinfo{person}{Raimund Dachselt}.}
  \bibinfo{year}{2020}\natexlab{}.
\newblock \showarticletitle{Augmented displays: Seamlessly extending
  interactive surfaces with head-mounted augmented reality}. In
  \bibinfo{booktitle}{\emph{Extended Abstracts of the 2020 CHI Conference on
  Human Factors in Computing Systems}}. \bibinfo{pages}{1--4}.
\newblock


\bibitem[Salthouse(1994)]%
        {salthouse1994}
\bibfield{author}{\bibinfo{person}{T.A. Salthouse}.}
  \bibinfo{year}{1994}\natexlab{}.
\newblock \showarticletitle{The aging of working memory}.
\newblock \bibinfo{journal}{\emph{Neuropsychology}}  \bibinfo{volume}{8}
  (\bibinfo{year}{1994}), \bibinfo{pages}{535--543}.
\newblock


\bibitem[Santos et~al\mbox{.}(2014)]%
        {Santos2014Augmented}
\bibfield{author}{\bibinfo{person}{Marc Ericson~C. Santos},
  \bibinfo{person}{Angie Chen}, \bibinfo{person}{Takafumi Taketomi},
  \bibinfo{person}{Goshiro Yamamoto}, \bibinfo{person}{Jun Miyazaki}, {and}
  \bibinfo{person}{Hirokazu Kato}.} \bibinfo{year}{2014}\natexlab{}.
\newblock \showarticletitle{Augmented Reality Learning Experiences: Survey of
  Prototype Design and Evaluation}.
\newblock \bibinfo{journal}{\emph{IEEE Transactions on Learning Technologies}}
  \bibinfo{volume}{7}, \bibinfo{number}{1} (\bibinfo{year}{2014}),
  \bibinfo{pages}{38--56}.
\newblock
\urldef\tempurl%
\url{https://doi.org/10.1109/TLT.2013.37}
\showDOI{\tempurl}


\bibitem[Schlagowski et~al\mbox{.}(2023)]%
        {Schlagowski2023Wish}
\bibfield{author}{\bibinfo{person}{Ruben Schlagowski}, \bibinfo{person}{Dariia
  Nazarenko}, \bibinfo{person}{Yekta Can}, \bibinfo{person}{Kunal Gupta},
  \bibinfo{person}{Silvan Mertes}, \bibinfo{person}{Mark Billinghurst}, {and}
  \bibinfo{person}{Elisabeth Andr\'{e}}.} \bibinfo{year}{2023}\natexlab{}.
\newblock \showarticletitle{Wish You Were Here: Mental and Physiological
  Effects of Remote Music Collaboration in Mixed Reality}. In
  \bibinfo{booktitle}{\emph{Proceedings of the 2023 CHI Conference on Human
  Factors in Computing Systems}} (Hamburg, Germany) \emph{(\bibinfo{series}{CHI
  '23})}. \bibinfo{publisher}{Association for Computing Machinery},
  \bibinfo{address}{New York, NY, USA}, Article \bibinfo{articleno}{102},
  \bibinfo{numpages}{16}~pages.
\newblock
\showISBNx{9781450394215}
\urldef\tempurl%
\url{https://doi.org/10.1145/3544548.3581162}
\showDOI{\tempurl}


\bibitem[Seidler et~al\mbox{.}(2010)]%
        {seidler2010}
\bibfield{author}{\bibinfo{person}{R.~D. Seidler}, \bibinfo{person}{J.~A.
  Bernard}, \bibinfo{person}{T.~B. Burutolu}, \bibinfo{person}{B.~W. Fling},
  \bibinfo{person}{M.~T. Gordon}, \bibinfo{person}{J.~T. Gwin},
  \bibinfo{person}{Y. Kwak}, {and} \bibinfo{person}{D.~B. Lipps}.}
  \bibinfo{year}{2010}\natexlab{}.
\newblock \showarticletitle{Motor control and aging: links to age-related brain
  structural, functional, and biochemical effects}.
\newblock \bibinfo{journal}{\emph{Neuroscience and Biobehavioral Reviews}}
  \bibinfo{volume}{34}, \bibinfo{number}{5} (\bibinfo{date}{Apr}
  \bibinfo{year}{2010}), \bibinfo{pages}{721--733}.
\newblock
\urldef\tempurl%
\url{https://doi.org/10.1016/j.neubiorev.2009.10.005}
\showDOI{\tempurl}
\newblock
\shownote{Epub 2009 Oct 20. PMID: 19850077. PMCID: PMC2838968}.


\bibitem[Shao et~al\mbox{.}(2023)]%
        {shao2023smartphone}
\bibfield{author}{\bibinfo{person}{Y. Shao}, \bibinfo{person}{J. Zhou}, {and}
  \bibinfo{person}{W. Wang}.} \bibinfo{year}{2023}\natexlab{}.
\newblock \showarticletitle{Smartphone touch gesture for right-handed older
  adults: touch performance and offset models}.
\newblock \bibinfo{journal}{\emph{J Ambient Intell Human Comput}}
  \bibinfo{volume}{14} (\bibinfo{year}{2023}), \bibinfo{pages}{2549--2566}.
\newblock
\urldef\tempurl%
\url{https://doi.org/10.1007/s12652-022-04502-8}
\showDOI{\tempurl}


\bibitem[Shareef et~al\mbox{.}(2011)]%
        {SHAREEF201117}
\bibfield{author}{\bibinfo{person}{Mahmud~Akhter Shareef},
  \bibinfo{person}{Vinod Kumar}, \bibinfo{person}{Uma Kumar}, {and}
  \bibinfo{person}{Yogesh~K. Dwivedi}.} \bibinfo{year}{2011}\natexlab{}.
\newblock \showarticletitle{e-Government Adoption Model (GAM): Differing
  service maturity levels}.
\newblock \bibinfo{journal}{\emph{Government Information Quarterly}}
  \bibinfo{volume}{28}, \bibinfo{number}{1} (\bibinfo{year}{2011}),
  \bibinfo{pages}{17--35}.
\newblock
\showISSN{0740-624X}
\urldef\tempurl%
\url{https://doi.org/10.1016/j.giq.2010.05.006}
\showDOI{\tempurl}


\bibitem[Sims et~al\mbox{.}(2017)]%
        {sims2017information}
\bibfield{author}{\bibinfo{person}{Tamara Sims}, \bibinfo{person}{Andrew~E
  Reed}, {and} \bibinfo{person}{Dawn~C Carr}.} \bibinfo{year}{2017}\natexlab{}.
\newblock \showarticletitle{Information and communication technology use is
  related to higher well-being among the oldest-old}.
\newblock \bibinfo{journal}{\emph{The Journals of Gerontology: Series B}}
  \bibinfo{volume}{72}, \bibinfo{number}{5} (\bibinfo{year}{2017}),
  \bibinfo{pages}{761--770}.
\newblock


\bibitem[Sinha and Swearingen(2013)]%
        {sinha2013}
\bibfield{author}{\bibinfo{person}{N. Sinha} {and} \bibinfo{person}{K.
  Swearingen}.} \bibinfo{year}{2013}\natexlab{}.
\newblock \showarticletitle{The role of cognitive decline in a digital age}.
\newblock \bibinfo{journal}{\emph{Gerontechnology}}  \bibinfo{volume}{11}
  (\bibinfo{year}{2013}), \bibinfo{pages}{536--544}.
\newblock


\bibitem[Suzuki et~al\mbox{.}(2022)]%
        {suzuki2022augmented}
\bibfield{author}{\bibinfo{person}{Ryo Suzuki}, \bibinfo{person}{Adnan Karim},
  \bibinfo{person}{Tian Xia}, \bibinfo{person}{Hooman Hedayati}, {and}
  \bibinfo{person}{Nicolai Marquardt}.} \bibinfo{year}{2022}\natexlab{}.
\newblock \showarticletitle{Augmented reality and robotics: A survey and
  taxonomy for ar-enhanced human-robot interaction and robotic interfaces}. In
  \bibinfo{booktitle}{\emph{Proceedings of the 2022 CHI Conference on Human
  Factors in Computing Systems}}. \bibinfo{pages}{1--33}.
\newblock


\bibitem[Tam and Chui(2016)]%
        {Tam2016Ageing}
\bibfield{author}{\bibinfo{person}{M. Tam} {and} \bibinfo{person}{E. Chui}.}
  \bibinfo{year}{2016}\natexlab{}.
\newblock \showarticletitle{Ageing and learning: What do they mean to elders
  themselves?}
\newblock \bibinfo{journal}{\emph{Studies in Continuing Education}}
  \bibinfo{volume}{38}, \bibinfo{number}{2} (\bibinfo{year}{2016}),
  \bibinfo{pages}{195--212}.
\newblock
\urldef\tempurl%
\url{https://doi.org/10.1080/0158037X.2015.1061492}
\showDOI{\tempurl}


\bibitem[Van~Gerven et~al\mbox{.}(2003)]%
        {VanGerven2003}
\bibfield{author}{\bibinfo{person}{P.W. Van~Gerven}, \bibinfo{person}{F. Paas},
  \bibinfo{person}{J.J. Van~Merriënboer}, \bibinfo{person}{M. Hendriks}, {and}
  \bibinfo{person}{H.G. Schmidt}.} \bibinfo{year}{2003}\natexlab{}.
\newblock \showarticletitle{The efficiency of multimedia learning into old
  age}.
\newblock \bibinfo{journal}{\emph{British Journal of Educational Psychology}}
  \bibinfo{volume}{73}, \bibinfo{number}{4} (\bibinfo{date}{Dec}
  \bibinfo{year}{2003}), \bibinfo{pages}{489--505}.
\newblock
\urldef\tempurl%
\url{https://doi.org/10.1348/000709903322591208}
\showDOI{\tempurl}


\bibitem[Venkatesh et~al\mbox{.}(2003)]%
        {Venkatesh2003}
\bibfield{author}{\bibinfo{person}{V. Venkatesh}, \bibinfo{person}{M.~G.
  Morris}, \bibinfo{person}{G.~B. Davis}, {and} \bibinfo{person}{F.~D. Davis}.}
  \bibinfo{year}{2003}\natexlab{}.
\newblock \showarticletitle{User acceptance of information technology: Toward a
  unified view}.
\newblock \bibinfo{journal}{\emph{MIS Quarterly}} (\bibinfo{year}{2003}),
  \bibinfo{pages}{425--478}.
\newblock


\bibitem[Venkatesh et~al\mbox{.}(2012a)]%
        {Venkatesh2012}
\bibfield{author}{\bibinfo{person}{V. Venkatesh}, \bibinfo{person}{J.~Y.
  Thong}, {and} \bibinfo{person}{X. Xu}.} \bibinfo{year}{2012}\natexlab{a}.
\newblock \showarticletitle{Consumer acceptance and use of information
  technology: extending the unified theory of acceptance and use of
  technology}.
\newblock \bibinfo{journal}{\emph{MIS Quarterly}} \bibinfo{volume}{36},
  \bibinfo{number}{1} (\bibinfo{year}{2012}), \bibinfo{pages}{157--178}.
\newblock


\bibitem[Venkatesh et~al\mbox{.}(2012b)]%
        {Venkatesh2012UTAUT2}
\bibfield{author}{\bibinfo{person}{Viswanath Venkatesh}, \bibinfo{person}{James
  Y.~L. Thong}, {and} \bibinfo{person}{Xin Xu}.}
  \bibinfo{year}{2012}\natexlab{b}.
\newblock \showarticletitle{Consumer Acceptance and Use of Information
  Technology: Extending the Unified Theory of Acceptance and Use of
  Technology}.
\newblock \bibinfo{journal}{\emph{MIS Quarterly}} \bibinfo{volume}{36},
  \bibinfo{number}{1} (\bibinfo{year}{2012}), \bibinfo{pages}{157--178}.
\newblock
\showISSN{02767783}
\urldef\tempurl%
\url{http://www.jstor.org/stable/41410412}
\showURL{%
\tempurl}


\bibitem[Villanueva et~al\mbox{.}(2020)]%
        {Villanueva2020Meta}
\bibfield{author}{\bibinfo{person}{Ana Villanueva}, \bibinfo{person}{Zhengzhe
  Zhu}, \bibinfo{person}{Ziyi Liu}, \bibinfo{person}{Kylie Peppler},
  \bibinfo{person}{Thomas Redick}, {and} \bibinfo{person}{Karthik Ramani}.}
  \bibinfo{year}{2020}\natexlab{}.
\newblock \showarticletitle{Meta-AR-App: An Authoring Platform for
  Collaborative Augmented Reality in STEM Classrooms}. In
  \bibinfo{booktitle}{\emph{Proceedings of the 2020 CHI Conference on Human
  Factors in Computing Systems}} (Honolulu, HI, USA)
  \emph{(\bibinfo{series}{CHI '20})}. \bibinfo{publisher}{Association for
  Computing Machinery}, \bibinfo{address}{New York, NY, USA},
  \bibinfo{pages}{1–14}.
\newblock
\showISBNx{9781450367080}
\urldef\tempurl%
\url{https://doi.org/10.1145/3313831.3376146}
\showDOI{\tempurl}


\bibitem[Villanueva et~al\mbox{.}(2021)]%
        {villanueva2021robotar}
\bibfield{author}{\bibinfo{person}{Ana~M Villanueva}, \bibinfo{person}{Ziyi
  Liu}, \bibinfo{person}{Zhengzhe Zhu}, \bibinfo{person}{Xin Du},
  \bibinfo{person}{Joey Huang}, \bibinfo{person}{Kylie~A Peppler}, {and}
  \bibinfo{person}{Karthik Ramani}.} \bibinfo{year}{2021}\natexlab{}.
\newblock \showarticletitle{Robotar: An augmented reality compatible
  teleconsulting robotics toolkit for augmented makerspace experiences}. In
  \bibinfo{booktitle}{\emph{Proceedings of the 2021 CHI Conference on Human
  Factors in Computing Systems}}. \bibinfo{pages}{1--13}.
\newblock


\bibitem[Wei et~al\mbox{.}(2023)]%
        {Xiaoying2023Bridging}
\bibfield{author}{\bibinfo{person}{Xiaoying Wei}, \bibinfo{person}{Yizheng Gu},
  \bibinfo{person}{Emily Kuang}, \bibinfo{person}{Xian Wang},
  \bibinfo{person}{Beiyan Cao}, \bibinfo{person}{Xiaofu Jin}, {and}
  \bibinfo{person}{Mingming Fan}.} \bibinfo{year}{2023}\natexlab{}.
\newblock \showarticletitle{Bridging the Generational Gap: Exploring How
  Virtual Reality Supports Remote Communication Between Grandparents and
  Grandchildren}. In \bibinfo{booktitle}{\emph{Proceedings of the 2023 CHI
  Conference on Human Factors in Computing Systems}} (Hamburg, Germany)
  \emph{(\bibinfo{series}{CHI '23})}. \bibinfo{publisher}{Association for
  Computing Machinery}, \bibinfo{address}{New York, NY, USA}, Article
  \bibinfo{articleno}{444}, \bibinfo{numpages}{15}~pages.
\newblock
\showISBNx{9781450394215}
\urldef\tempurl%
\url{https://doi.org/10.1145/3544548.3581405}
\showDOI{\tempurl}


\bibitem[Williams et~al\mbox{.}(2021)]%
        {Williams2021Augmented}
\bibfield{author}{\bibinfo{person}{Thomas~J. Williams},
  \bibinfo{person}{Simon~L. Jones}, \bibinfo{person}{Christof Lutteroth},
  \bibinfo{person}{Elies Dekoninck}, {and} \bibinfo{person}{Hazel~C Boyd}.}
  \bibinfo{year}{2021}\natexlab{}.
\newblock \showarticletitle{Augmented Reality and Older Adults: A Comparison of
  Prompting Types}. In \bibinfo{booktitle}{\emph{Proceedings of the 2021 CHI
  Conference on Human Factors in Computing Systems}} (Yokohama, Japan)
  \emph{(\bibinfo{series}{CHI '21})}. \bibinfo{publisher}{Association for
  Computing Machinery}, \bibinfo{address}{New York, NY, USA}, Article
  \bibinfo{articleno}{723}, \bibinfo{numpages}{13}~pages.
\newblock
\showISBNx{9781450380966}
\urldef\tempurl%
\url{https://doi.org/10.1145/3411764.3445476}
\showDOI{\tempurl}


\bibitem[Wilson et~al\mbox{.}(2018)]%
        {Wilson2018Help}
\bibfield{author}{\bibinfo{person}{Zachary Wilson}, \bibinfo{person}{Helen
  Yin}, \bibinfo{person}{Sayan Sarcar}, \bibinfo{person}{Rock Leung}, {and}
  \bibinfo{person}{Joanna McGrenere}.} \bibinfo{year}{2018}\natexlab{}.
\newblock \showarticletitle{Help Kiosk: An Augmented Display System to Assist
  Older Adults to Learn How to Use Smart Phones}. In
  \bibinfo{booktitle}{\emph{Proceedings of the 20th International ACM SIGACCESS
  Conference on Computers and Accessibility}} (Galway, Ireland)
  \emph{(\bibinfo{series}{ASSETS '18})}. \bibinfo{publisher}{Association for
  Computing Machinery}, \bibinfo{address}{New York, NY, USA},
  \bibinfo{pages}{441–443}.
\newblock
\showISBNx{9781450356503}
\urldef\tempurl%
\url{https://doi.org/10.1145/3234695.3241008}
\showDOI{\tempurl}


\bibitem[Xiao et~al\mbox{.}(2018)]%
        {xiao2018mrtouch}
\bibfield{author}{\bibinfo{person}{Robert Xiao}, \bibinfo{person}{Julia
  Schwarz}, \bibinfo{person}{Nick Throm}, \bibinfo{person}{Andrew~D Wilson},
  {and} \bibinfo{person}{Hrvoje Benko}.} \bibinfo{year}{2018}\natexlab{}.
\newblock \showarticletitle{MRTouch: Adding touch input to head-mounted mixed
  reality}.
\newblock \bibinfo{journal}{\emph{IEEE transactions on visualization and
  computer graphics}} \bibinfo{volume}{24}, \bibinfo{number}{4}
  (\bibinfo{year}{2018}), \bibinfo{pages}{1653--1660}.
\newblock


\bibitem[Yusif et~al\mbox{.}(2016)]%
        {Yusif2016Older}
\bibfield{author}{\bibinfo{person}{Salifu Yusif}, \bibinfo{person}{Jeffrey
  Soar}, {and} \bibinfo{person}{Abdul Hafeez-Baig}.}
  \bibinfo{year}{2016}\natexlab{}.
\newblock \showarticletitle{Older people, assistive technologies, and the
  barriers to adoption: A systematic review}.
\newblock \bibinfo{journal}{\emph{International Journal of Medical
  Informatics}}  \bibinfo{volume}{94} (\bibinfo{date}{07}
  \bibinfo{year}{2016}).
\newblock
\urldef\tempurl%
\url{https://doi.org/10.1016/j.ijmedinf.2016.07.004}
\showDOI{\tempurl}


\bibitem[Zhu and Grossman(2020)]%
        {zhu2020bishare}
\bibfield{author}{\bibinfo{person}{Fengyuan Zhu} {and} \bibinfo{person}{Tovi
  Grossman}.} \bibinfo{year}{2020}\natexlab{}.
\newblock \showarticletitle{Bishare: Exploring bidirectional interactions
  between smartphones and head-mounted augmented reality}. In
  \bibinfo{booktitle}{\emph{Proceedings of the 2020 CHI Conference on Human
  Factors in Computing Systems}}. \bibinfo{pages}{1--14}.
\newblock


\bibitem[Ziefle(2010)]%
        {ZIEFLE2010719}
\bibfield{author}{\bibinfo{person}{Martina Ziefle}.}
  \bibinfo{year}{2010}\natexlab{}.
\newblock \showarticletitle{Information presentation in small screen devices:
  The trade-off between visual density and menu foresight}.
\newblock \bibinfo{journal}{\emph{Applied Ergonomics}} \bibinfo{volume}{41},
  \bibinfo{number}{6} (\bibinfo{year}{2010}), \bibinfo{pages}{719--730}.
\newblock
\showISSN{0003-6870}
\urldef\tempurl%
\url{https://doi.org/10.1016/j.apergo.2010.03.001}
\showDOI{\tempurl}
\newblock
\shownote{Special Section: Selection of papers from IEA 2009}.


\bibitem[Åberg(2016)]%
        {Aberg2016Nonformal}
\bibfield{author}{\bibinfo{person}{P. Åberg}.}
  \bibinfo{year}{2016}\natexlab{}.
\newblock \showarticletitle{Nonformal learning and well-being among older
  adults: Links between participation in Swedish study circles, feelings of
  well-being and social aspects of learning}.
\newblock \bibinfo{journal}{\emph{Educational Gerontology}}
  \bibinfo{volume}{42}, \bibinfo{number}{6} (\bibinfo{year}{2016}),
  \bibinfo{pages}{411--422}.
\newblock
\urldef\tempurl%
\url{https://doi.org/10.1080/03601277.2016.1139972}
\showDOI{\tempurl}


\end{thebibliography}
\end{document}